\documentclass[usenatbib]{mn2e}
\usepackage{graphicx,times, amsmath, epsfig}

\newcommand{\taure}{\tau_{\rm reion}}
\newcommand{\tauHII}{\tau_{\rm HII}}
\newcommand{\Dss}{\Delta_{\rm ss}}

\newcommand{\GammaHII}{\Gamma_{\rm HII}}
\newcommand{\aveGammaHII}{\langle \Gamma_{12} \rangle_{\rm HII}}
\newcommand{\avenfHII}{\langle x_{\rm HI} \rangle_{\rm HII}}

\newcommand{\QHII}{Q_{\rm HII}}

\newcommand{\dexm}{\textsc{\small DexM}}
\newcommand{\enzo}{\textsc{\small ENZO}}

\newcommand{\cmfast}{\textsc{\small 21CMFAST}}

\newcommand{\nf}{x_{\rm HI}}

\newcommand{\lya}{Ly$\alpha$}

\newcommand{\Msun}{M_\odot}

\newcommand{\Mmin}{M_{\rm min}}

\newcommand{\lmfp}{\lambda_{\rm mfp}}

\newcommand\lsim{\mathrel{\rlap{\lower4pt\hbox{\hskip1pt$\sim$}}
        \raise1pt\hbox{$<$}}}
\newcommand\gsim{\mathrel{\rlap{\lower4pt\hbox{\hskip1pt$\sim$}}
        \raise1pt\hbox{$>$}}}
\def\myputfigure#1#2#3#4#5%
{\vskip#5pt\makebox[0pt]{\hskip#2in
\includegraphics[width=#3\textwidth]{#1}}\vskip#4pt\hfill}

\pdfoutput=1

\begin{document}

\title[Can the IGM cause a rapid drop in Ly$\alpha$ emission at $z>6$?]{Can the intergalactic medium cause a rapid drop in Ly$\alpha$ emission at $z>6$?}

\author[Mesinger et al.]{Andrei Mesinger$^1$\thanks{email: andrei.mesinger@sns.it},
Aycin Aykutalp$^1$,
Eros Vanzella$^2$,
Laura Pentericci$^3$,
Andrea Ferrara$^1$,
\newauthor ~ \& Mark Dijkstra$^4$\\
$^1$Scuola Normale Superiore, Piazza dei Cavalieri 7, 56126 Pisa, Italy\\
$^2$INAF Osservatorio Astronomico di Bologna, Italy\\
$^3$INAF Osservatorio Astronomico di Roma, Via Frascatti 33, 00040 Monteporzio, RM, Italy\\
$^4$Institute of Theoretical Astrophysics, University of Oslo, Postboks 1029, 0858 Oslo, Norway
}

\voffset-.6in

\maketitle

\begin{abstract}
The large cross-section of the \lya\ line makes it a sensitive probe of the ionization state of the intergalactic medium (IGM). 
  Here we present the most complete study to date of the IGM \lya\ opacity, and its application to the redshift evolution of the '\lya\ fraction',  i.e. the fraction of color-selected galaxies with a detectable \lya\ emission line.  We use a tiered approach, which combines large-scale semi-numeric simulations of reionization with moderate-scale hydrodynamic simulations of the ionized IGM.  This allows us to simultaneously account for evolution in both: (i) the opacity from an incomplete (patchy) reionization, parameterized by the filling factor of ionized regions, $Q_{\rm HII}$; and (ii) the opacity from self-shielded systems in the ionized IGM, parameterized by the average photo-ionization rate inside HII regions, $\aveGammaHII$.
In contrast to recent empirical models, attenuation from patchy reionization has a unimodal distribution along different sightlines, while attenuation from self-shielded systems is more bimodal.
We quantify the average IGM transmission in our ($Q_{\rm HII}$, $\aveGammaHII$) parameter space, which can easily be used to interpret new data sets. 
Using current observations, we predict that the \lya\ fraction cannot drop by more than a factor of $\approx2$ with IGM attenuation alone, even for HII filling factors as low as $Q_{\rm HII}\gsim$0.1. Larger changes in the \lya\ fraction could result from a co-evolution with galaxy properties.
 Marginalizing over $\aveGammaHII$, 
 we find that current observations constrain $Q_{\rm HII}(z\approx7) \leq 0.6$, at a 68\% confidence level (C.L.).  However, all of our parameter space is consistent with observations at 95\% C.L., highlighting the need for larger observational samples at $z\geq6$. 
\end{abstract}


\section{Introduction}
\label{sec:intro}

The Lyman alpha line from galaxies provides a window on the last frontier in astrophysical cosmology: the Epoch of Reionization (EoR).  \lya\ is strongly attenuated by neutral hydrogen, even in the damping wing of the line cross-section.  Neutral patches of the intergalactic medium (IGM) could therefore significantly suppress the intrinsic \lya\ line of galaxies during reionization (e.g. \citealt{Miralda-Escude98, HS99, Santos04}).

Extracting the imprint of reionization from galaxy observations at a single redshift is difficult, since we do not know how to apriori model the intrinsic \lya\ line shape and luminosity, before it is processed by IGM absorption (see \S \ref{sec:line}).  Therefore, reionization constraints generally rely on the redshift {\it evolution} of \lya\ observables, which allows one to assume that galaxy and line properties do not evolve over the same interval.  Indeed, recently it was claimed that the fraction of color-selected galaxies with a strong \lya\ line (generally defined as having a rest frame equivalent width greater than 25 \AA), the so-called \lya\ fraction, drops sharply from $z\approx6\rightarrow7$; although interpretation is still limited by small number statistics (e.g. \citealt{Stark10, Pentericci11, Ono12, Schenker14}).  Absorption by neutral patches remaining from an incomplete reionization is often evoked as an obvious cause of such a drop.

However, the late stages of reionization are characterized by large cosmic HII regions.  The remaining neutral patches are generally distant from galaxies, with a correspondingly weak damping wing imprint on the \lya\ line.  A strong evolution in the \lya\ fraction would therefore require a very substantial change in the filling factor of ionized regions, $Q_{\rm HII}$, over the same brief interval: $z\approx6\rightarrow7$ (e.g. \citealt{McQuinn07LAE, MF08LAE, DMW11, Jensen13}).  Alternately, if the photo-ionizing background drops rapidly beyond $z\gsim6$, the increasing abundance of self-shielded systems inside the ionized IGM can imprint a similar signature \citep{BH13}.  Yet another possibility to explain a drop in the \lya\ fraction is an evolution of the intrinsic galaxy properties themselves, such as wind characteristics and the escape fraction of ionizing photons, $f_{\rm esc}$ \citep{Dijkstra14}.  Alternately, a joint evolution in the IGM and/or galaxy properties could ease tension with observations.


Here we develop the most comprehensive model of IGM absorption to date, including the impact of both (i) the large-scale (hundreds of cMpc; e.g. \citealt{Iliev14}) reionization morphology; and (ii) $\sim$kpc scale (e.g. \citealt{Schaye01}) self-shielded systems.  We make use of well-tested semi-numerical simulations to model reionization morphology.  Our reionization simulations include sub-grid modeling of UV photo-heating feedback and recombinations, shown to significantly decrease the size of cosmic HII regions \citep{SM14}. We complement these with moderate-scale (tens of cMpc) hydrodynamic simulations of the ionized IGM, resolving the relevant self-shielded systems.  With this tiered approach, we show how {\it the redshift evolution in the $z\gsim6$ \lya\ fraction can constrain both the filling factor of ionized regions as well as the photo-ionizing background.}

This paper is organized as follows. In \S \ref{sec:model}, we present our model for IGM absorption, sourced by both patchy reionization (\S \ref{sec:reion}) and self-shielded systems (\S \ref{sec:IGM}). In \S \ref{sec:results}, we present our results on the evolution of the IGM transmission and the \lya\ fraction from $z\approx6\rightarrow7$.  Finally, we conclude in \S \ref{sec:conc}.

Unless stated otherwise, we quote all quantities in comoving units. We adopt the background cosmological parameters: ($\Omega_\Lambda$, $\Omega_{\rm M}$, $\Omega_b$, $n$, $\sigma_8$, $H_0$) = (0.68, 0.32, 0.049, 0.96, 0.83, 67 km s$^{-1}$ Mpc$^{-1}$), consistent with recent results from the Planck mission \citep{Planck13}.

\section{Model}
\label{sec:model}

\begin{figure*}
\vspace{-1\baselineskip}
{
\includegraphics[width=\textwidth]{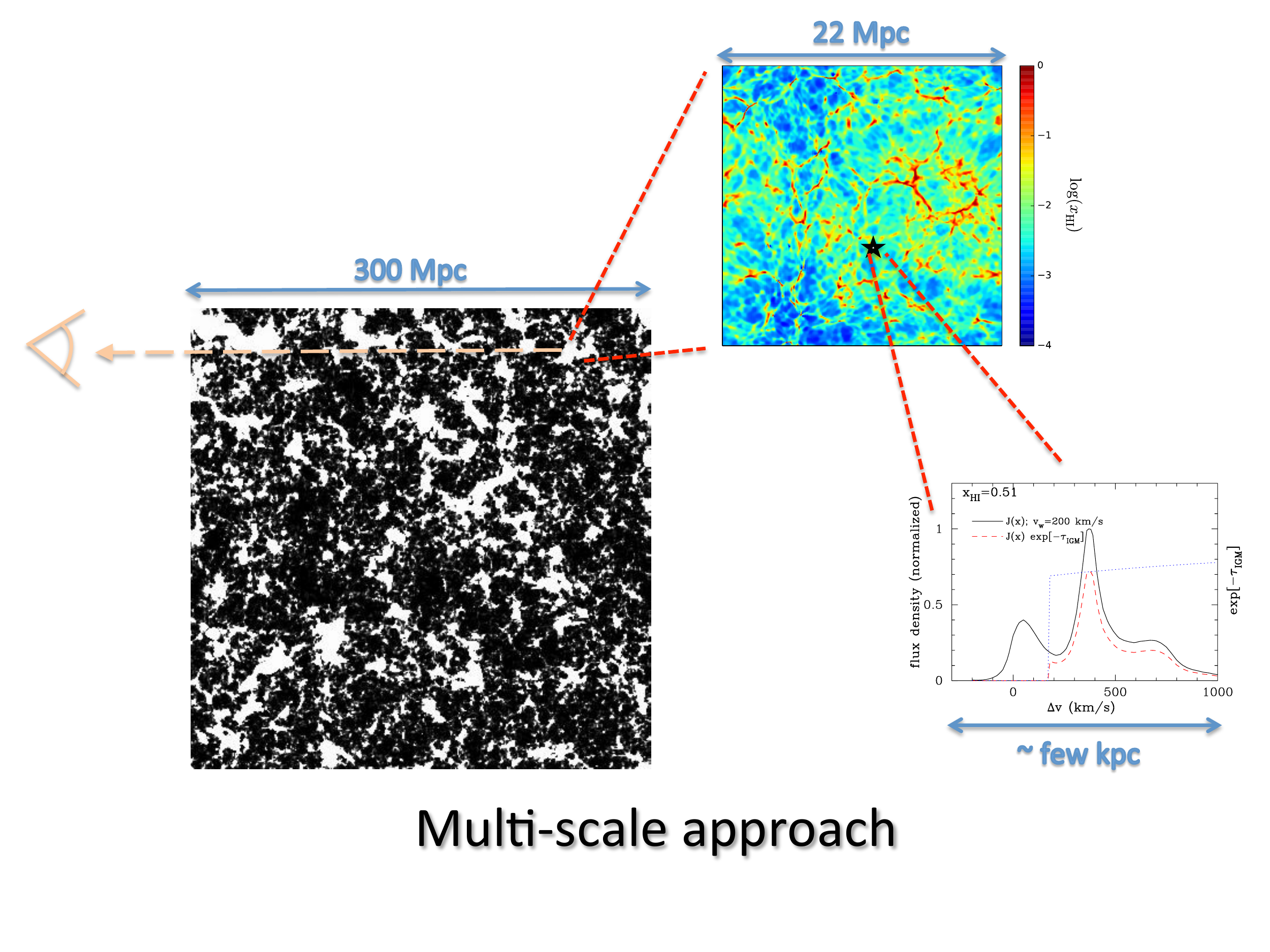}
}
\vspace{-2\baselineskip}
\caption{
Schematic showing the various components of our model.  From left to right we show: (i) a 0.75 Mpc thick slice through our large-scale reionization simulation at $\QHII \sim 0.5$ \citep{SM14}; (ii) a 21 kpc slice through our hydro simulation of the ionized IGM surrounding high-$z$ galaxies; (iii) the \lya\ line emerging from a galaxy including RT through local outflows.  (i) and (ii) are used in this work, while (iii) is taken from \citet{DMW11}. Relative scales are approximate.
}
\label{fig:schematic}
\vspace{-0.5\baselineskip}
\end{figure*}

Because of the difficulties mentioned above, we model the total opacity using a tiered approach, illustrated in Fig. \ref{fig:schematic}.  We simulate the morphology of reionization on large scales (hundreds of Mpc), using a semi-numeric simulation.  We complement this with a moderate-scale (tens of Mpc) hydrodynamic simulation, that resolves the high-column density systems inside the cosmic HII regions.

In other words, we split the total optical depth into a component sourced by the reionization morphology, $\taure$, and from the {\it residual} neutral hydrogen {\it inside} the local HII regions surrounding galaxies, $\tauHII$, so that the total \lya\ optical depth is:
\begin{align}
\label{eq:tautot}
\tau(Q_{\rm HII}, \aveGammaHII) &= \taure(Q_{\rm HII}) + \tauHII(\aveGammaHII) \\
\nonumber &= \int_{R_{\rm HII}}^{R_{\rm max}} d\taure + \int_{R_{\rm min}}^{R_{\rm HII}} d\tauHII ~ .
\end{align}
Here, $Q_{\rm HII}$ is the filling factor of HII regions and $\aveGammaHII$ is the spatially-averaged photo-ionization rate (in units of 10$^{-12}$ s$^{-1}$) {\it inside} HII regions\footnote{This formalism assumes that the photo-ionization rate inside neutral IGM patches, $\langle \Gamma_{12} \rangle_{\rm HI}$, is negligible (i.e. that the patches are fully neutral), consistent with stellar-driven reionization models (see, e.g., appendix of \citealt{Zahn11}).  Thus the total,  volume-averaged photo-ionization rate can be approximated as: $\langle \Gamma_{12} \rangle_{\rm V} \approx (1-Q_{\rm HII}) \langle \Gamma_{12} \rangle_{\rm HI} + Q_{\rm HII} \aveGammaHII \approx Q_{\rm HII} \aveGammaHII$}.  {\it These are the two fundamental free parameters in our model.}\footnote{Note that $\aveGammaHII$ depends on the {\it instantaneous} ionizing emissivity and mean free path, while $Q_{\rm HII}$ depends (roughly) on the {\it time-integrated history} of these quantities. In principle, complete models of reionization could predict both $Q_{\rm HII}$ and $\aveGammaHII$ self-consistently (see the discussion in \S \ref{sec:gamma}).  However, as the relation between $Q_{\rm HII}$ and $\aveGammaHII$ is highly model-dependent and very uncertain, we leave these two quantities as free parameters.}

The sightlines used to compute $\taure$ and $\tauHII$ originate from independent halo catalogues from the two simulation boxes; we use 5000 randomly-oriented sightlines, originating from 100 halos with masses $M_{\rm halo}\approx 10^{10.5-11} \Msun$ (consistent with clustering measurements of \lya\ emitters; \citealt{Ouchi10}).  As indicated by eq. (\ref{eq:tautot}), each pair of sightlines through the two independent simulation boxes is traversed concurrently\footnote{Note that our approach neglects correlations between self-shielded systems and the reionization morphology.  This amounts to assuming that their correlation length is smaller than the typical HII region size, $\langle R_{\rm HII} \rangle$.  This is likely a safe assumption because: (i) self-shielded systems are only relevant for \lya\ if they are abundant, with correspondingly small correlation lengths; (ii) as we discuss further below, $z\sim7$ likely corresponds to the advanced stages of reionization when HII regions are very large, $\langle R_{\rm HII} \rangle \gsim$ tens of cMpc \citep{FZH04, Zahn11}.  In principle, suites of hydro simulations with varying matter densities could be used to account for these correlations.}.  $R_{\rm HII}$ corresponds to the distance from a halo to the edge of its surrounding cosmic HII region, following a randomly-oriented sightline through the large-scale reionization simulation.  For distances less than $R_{\rm HII}$, the sightline's opacity is fully determined by the hydrodynamic simulation ($\tauHII$), while at greater distances it is fully determined by the reionization simulation ($\taure$).
Hence,   $\taure$ is sourced by order unity fluctuations of the large-scale reionization field, while $\tauHII$ is sourced by the incidence rate of self-shielded, high-column density systems\footnote{Our formalism for $\tauHII$ accounts for the absorption by {\it all} of the gas inside the local HII region, not just from the self-shielded systems.} (see Fig. \ref{fig:schematic}).
 As we wish to avoid modeling the complexities of the interstellar medium (ISM) and circumgalactic medium (CGM), we start the optical depth integral at a (somewhat arbitrary) distance $R_{\rm min}$=0.16 cMpc away from the halo center, matching the choice in \citet{BH13}.  The ISM and CGM affect the line shape of the \lya\ line as it emerges from a galaxy. By varying the line shape, we address uncertainties associated with scattering process on these smaller scales. We compute the integrated opacity of material extending out to a total distance of $R_{\rm max}=100$ cMpc.

 Finally, to illustrate the effects of the emerging \lya\ line profile, we assume a simple Gaussian profile, centered at various velocity offsets.  Modeling the intrinsic \lya\ emission is beyond the scope of this work.  However, our models illustrate the relevant trends, showing that the derived constraints are not very sensitive to the profile (though they could be sensitive to a redshift {\it evolution} in the profile).
Below we discuss the details of our model in turn.

\subsection{Reionization morphology}
\label{sec:reion}

\begin{figure*}
\vspace{-1\baselineskip}
{
\includegraphics[width=0.33\textwidth]{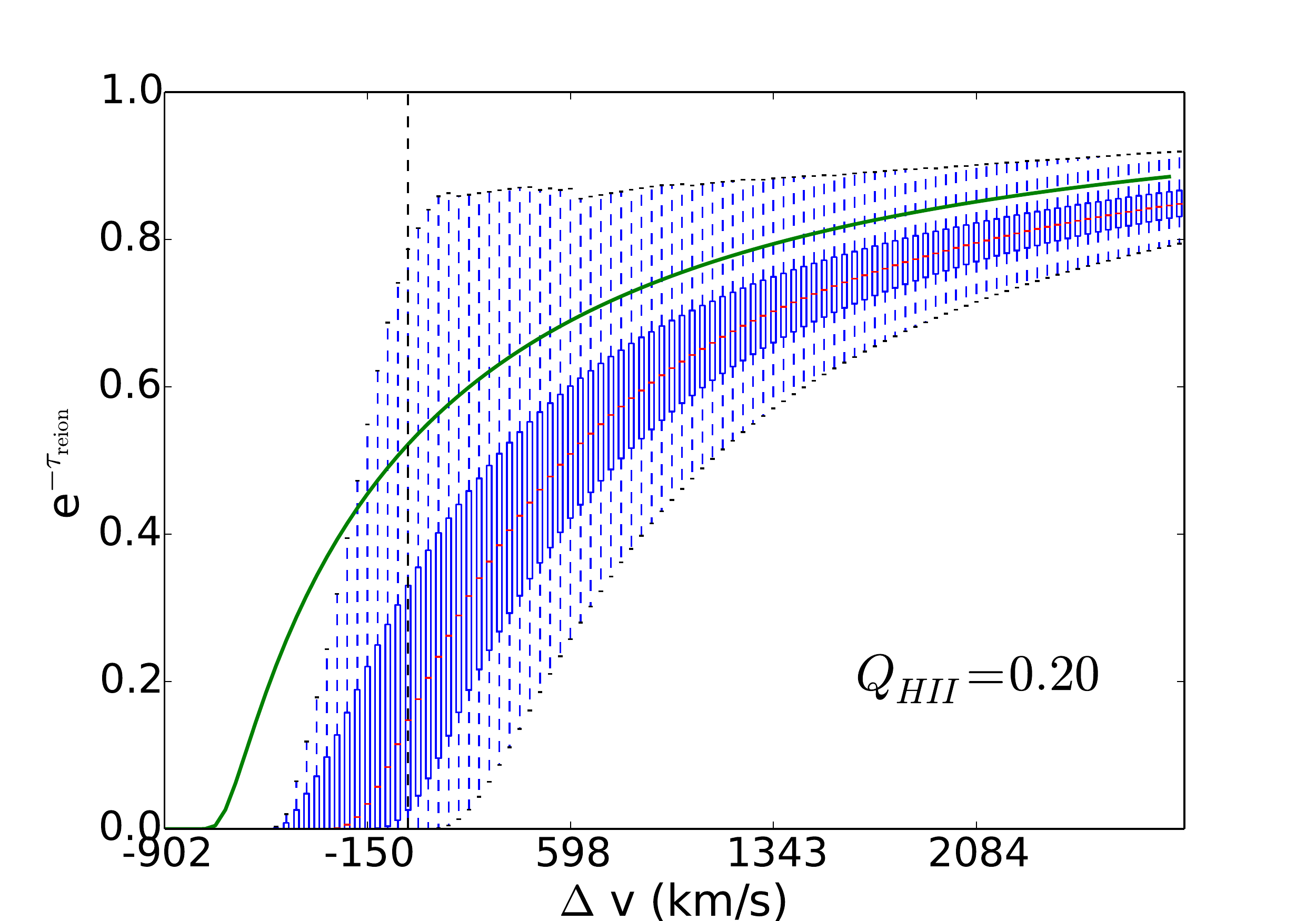}
\includegraphics[width=0.33\textwidth]{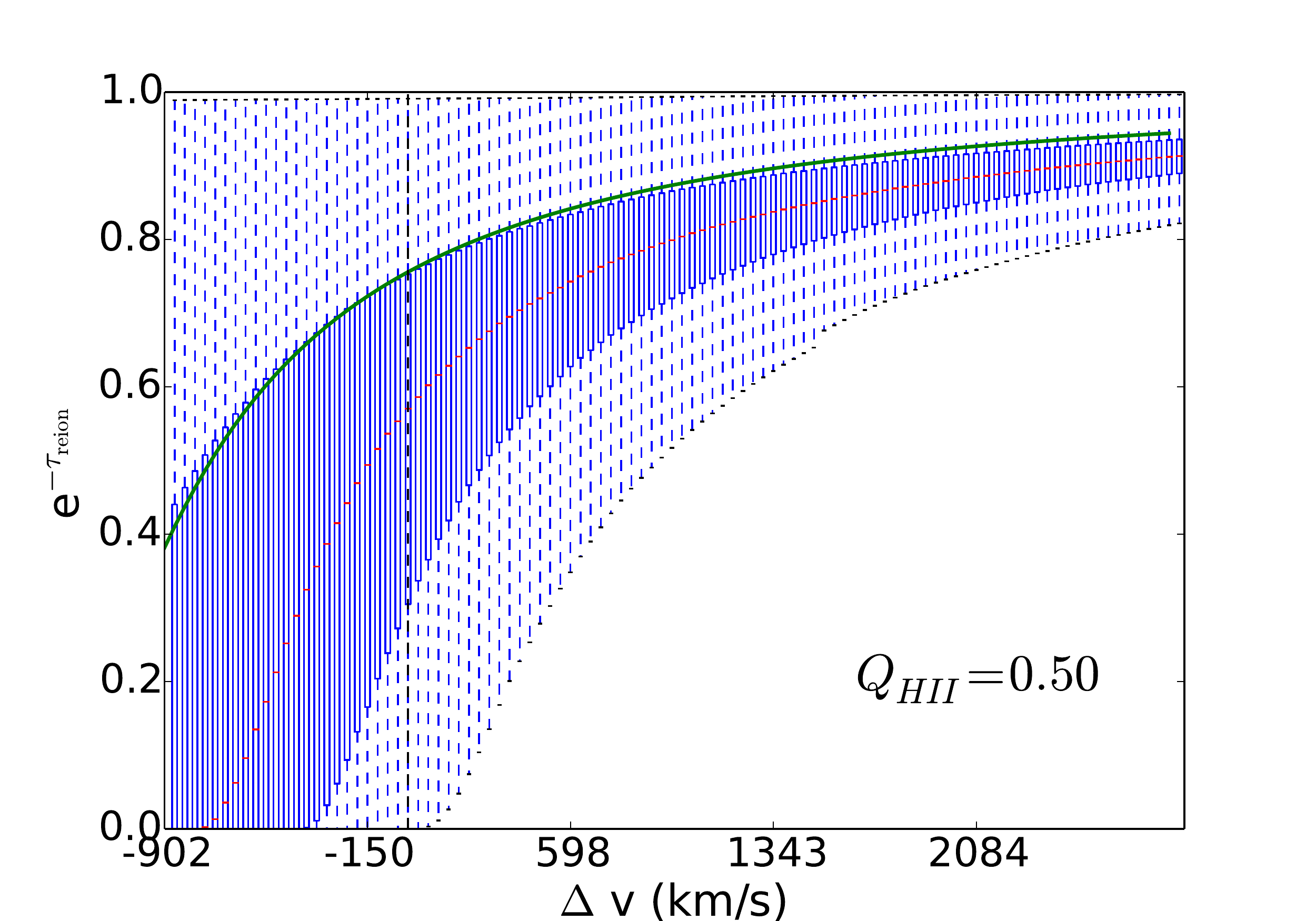}
\includegraphics[width=0.33\textwidth]{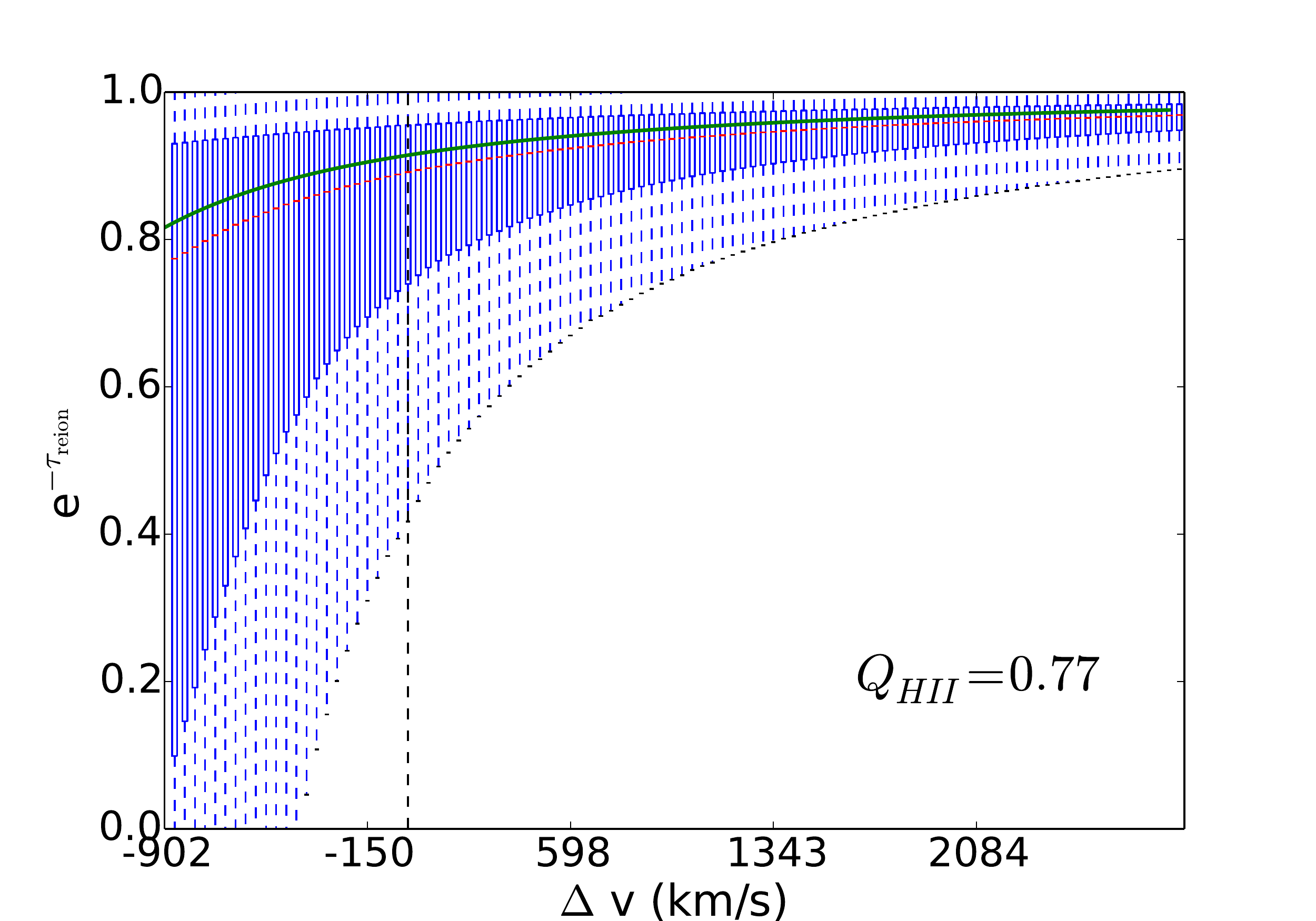}
}
\caption{
Distributions of the damping wing opacity from an inhomogeneous reionization, as functions of the rest-frame velocity offset from the \lya\ line center (shown as a vertical dashed line). Panels correspond to $\QHII=$ 0.20, 0.50, 0.77, ({\it left to right}).  Red lines correspond to the mean profile, while boxes/whiskers enclose the first two quartiles of the distribution.  The scatter represents the sightline-to-sightline scatter, but it should be noted that damping wings of individual profiles are smooth functions of wavelength.  Absorption profiles from a sample sightline are over-plotted with green curves.
}
\label{fig:taud}
\vspace{-0.5\baselineskip}
\end{figure*}

We model the large-scale EoR morphology using the publicly-available, semi-numerical codes \dexm\ and \cmfast\footnote{http://homepage.sns.it/mesinger/Sim} \citep{MF07, MFC11}.  The semi-numerical approach combines excursion set and perturbation formalisms to generate various cosmic fields, and has been extensively tested against numerical simulations \citep{MF07, MFC11, Zahn11}.

Our simulation box is 300 Mpc on a side with a resolution of 400$^3$.  For our ionization fields, we use the results from \citet{SM14}, which include inhomogeneous, sub-grid modeling of (i) UV feedback on galaxies; and (ii) recombinations.  Effect (i) results from the UVB heating the accreting gas, thereby suppressing the 
 star-formation rate in low-mass galaxies inside cosmic HII regions (e.g \citealt{Efstathiou92, SGB94, HG97}).  Using suites of spherically-symmetric cosmological hydro simulations, \citet{SM13a} provided a parameterized fit for the resulting depletion of gas and effective critical halo mass capable of hosting star-forming galaxies.
Likewise, (ii) accounts for inhomogeneous recombinations, which can be an important drain on the ionizing photon budget.  Both (i) and (ii) depend on the thermal and ionization history of the local gas patch, on scales which are too small to be tracked in reionization simulations, without a sub-grid prescription.  Furthermore, both effects are additive, in that they both preferentially impede the growth of large HII regions.  As both photo-heating feedback and recombinations act on non-negligible time-scales, these feedback effects are most pronounced in the biased centers of large HII regions, which were the first to ionize.

We generate halo catalogs at $z=7$ corresponding to the same initial conditions, using the excursion-set and perturbation theory approach outlined in \citet{MF07}. We generate 5000 randomly-oriented sightlines, originating from 100 halos with masses $M_{\rm halo}\approx 10^{10.5-11} \Msun$, roughly corresponding to the host halo masses of the observed population of high-$z$ galaxies, inferred through abundance matching and clustering (e.g. \citealt{Ouchi10}).  We compute $\taure$ by summing the damping wing contributions from large-scale, neutral patches of the mean density IGM (eq. 1 in \citealt{MF08LAE}), out to a comoving distance of 100 Mpc away from such halos.

It is important to note that this approach allows us to generate reionization fields from sources much fainter than the actual observed galaxies.  It is this population of faint undetected sources, residing in halos with masses $M_{\rm halo}\gsim 10^8$, which are expected to dominate reionization and govern its morphology (e.g. \citealt{CFG08, KF-G12, SM13b}).

 The resulting distributions of reionization opacities, $\exp[-\taure]$, are shown in Fig. \ref{fig:taud}, as functions of the velocity offset from the systemic redshift of the galaxy (shown as a vertical dashed line).  Panels correspond to $\QHII=$ 0.20, 0.50, 0.77, ({\it left to right}).

 During the advanced stages of reionization, most galaxies reside in large HII regions, suffering from only modest absorption from the remaining neutral patches.  This is further illustration that the observability of galactic \lya\ emission does not evolve dramatically during the end stages of reionization (e.g. \citealt{FZH06}).

Furthermore, the sightline-to-sightline scatter in the optical depth, $\taure$, rapidly increases as reionization progresses: some sightlines go through long stretches of the ionized IGM, while others encounter neutral material close to the galaxies.  Indeed, given enough sightlines, this scatter itself can be used as a probe of reionization \citep{MF08damp}.  Although the sightline-to-sightline scatter is large, individual sightlines have damping wing opacities which are smooth functions of wavelength (e.g. \citealt{Miralda-Escude98}).  One example sightline is illustrated with a green curve in the panels of Fig. \ref{fig:taud}.

\subsubsection{The role of IGM recombinations and UVB feedback in shaping reionization morphology}
\label{sec:recomb}

\begin{figure}
\vspace{-1\baselineskip}
{
\includegraphics[width=0.5\textwidth]{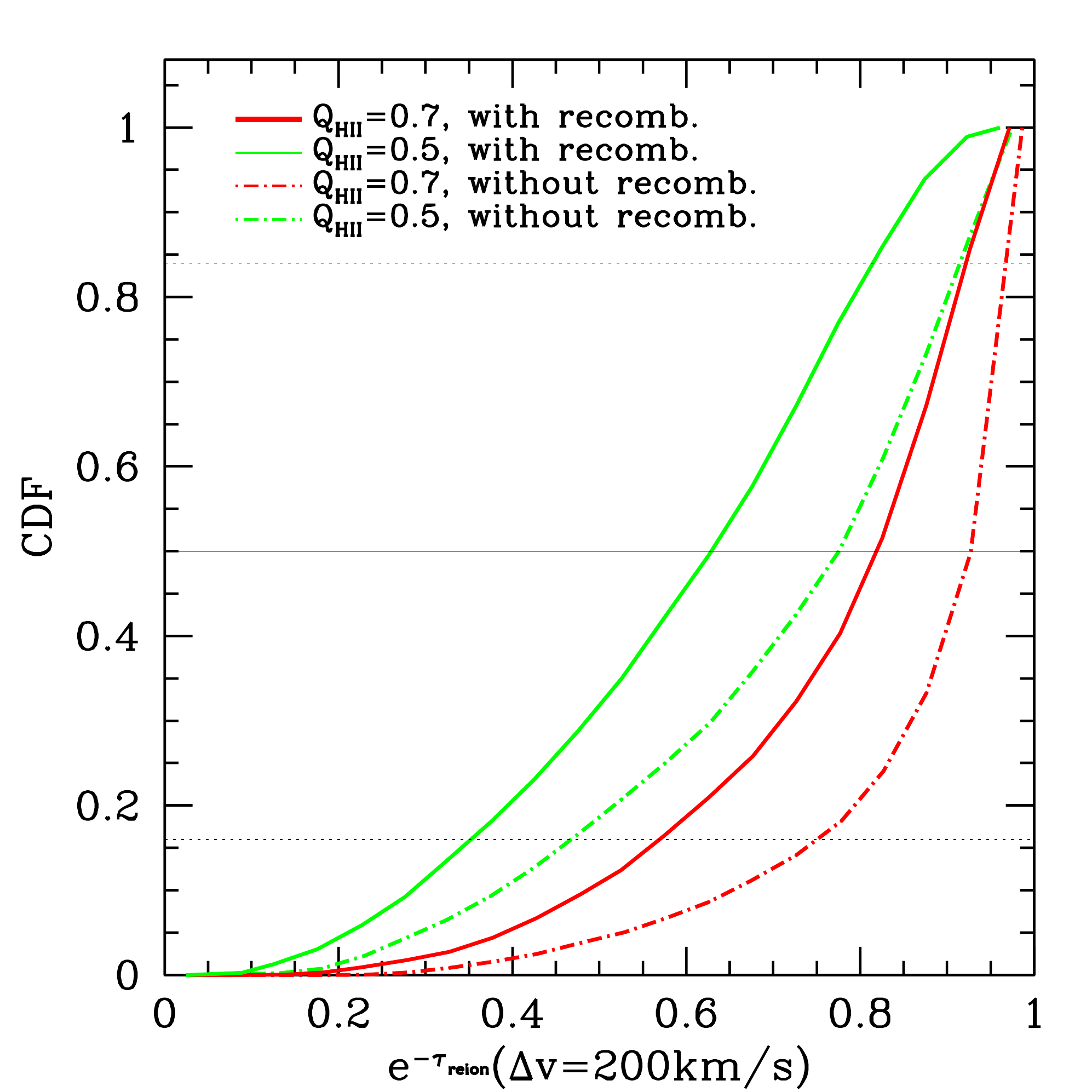}
}
\caption{
Fraction of sightlines having reionization opacities less than $\exp[-\taure]$, evaluated at $\Delta v$ = 200 km s$^{-1}$ redward of the galaxy's systemic redshift.
 The mean and 68\% contours are shown with thin horizontal lines. Curves correspond to $Q_{\rm HII} =$ 0.5, 0.7 ({\it left to right}).  The importance of self-consistently including IGM recombinations (and UVB feedback) is shown by comparing the solid to the dot-dashed curves.  The former corresponds to our fiducial reionization morphology (taken from \citealt{SM14}), while the later corresponds to the common approach of a redshift-independent minimum mass for star-forming galaxies (here taken to be $M_{\rm halo}=5\times10^8 \Msun$), and more importantly, ignoring the impact of inhomogeneous recombinations in the IGM.
}
\label{fig:reion_CDFs}
\vspace{-0.5\baselineskip}
\end{figure}

Using a sub-grid model of inhomogeneous recombinations, \citet{SM14} recently showed that recombinations in the IGM (driven by systems unresolved in reionization simulations) can dramatically suppress
 large-scale ionization structure\footnote{As discussed above, the full model of \citet{SM14} which we use for our morphologies also includes feedback by inhomogeneous reionization on the baryon content of reionizing galaxies.  Recombinations play a larger role than this UVB feedback in regulating the progress and morphology of reionization.  However, since both effects preferentially impact the largest HII regions, their combined impact is greater than their individual ones.}.
  At $\QHII\gsim0.9$, reionization progresses in a recombination-limited fashion, with recombinations balancing ionizations in large HII regions \citep{FO05}.  However, recombinations slow the growth of HII regions well before this cosmic Str{\"o}mgren limit, resulting in a more uniform reionization morphology with a dramatic supression of large HII regions.  Typical HII regions sizes are smaller by factors of $\sim$2--3 throughout reionization (Fig. 7 in \citealt{SM14}).  This ``feedback-limited''morphology is important because smaller HII regions increase the imprint of neutral cosmic gas on the \lya\ emission line.  Hence the change in $Q_{\rm HII}$ required to explain a given drop in the \lya\ fraction would be reduced, when compared to prior studies.

We show the impact of these feedback-limited reionization morphologies on the \lya\ transmission in Fig. \ref{fig:reion_CDFs}.  The CDFs are constructed from our 5000 sightlines, and evaluated at a single wavelength, $\Delta v$ = 200 km s$^{-1}$ redward of the systemic galaxy redshift. Solid curves correspond to our fiducial reionization morphologies, while the dot-dashed curves correspond to previous estimates which ignore the role of inhomogeneous recombinations and UV feedback in suppressing large HII regions.  From the figure, we see that our opacity distribution at $Q_{\rm HII}=0.7$ is similar to the one at an earlier stage in reionization, $Q_{\rm}\approx0.5-0.6$, using morphologies not taking into account recombinations and UVB feedback.  Hence, we already expect that {\it the \lya\ fraction constraints on $Q_{\rm HII}(z=7)$ will relax by $\Delta z \sim 0.1-0.2$, when taking into account new, feedback-limited morphologies.}


\subsection{Inside the ionized patches of the IGM}
\label{sec:IGM}

Even in the early stages of reionization, each galaxy is surrounded by a local HII region.  The residual volume-weighted neutral fraction inside these ionized patches is fairly modest (e.g. $\avenfHII \sim 10^{-3}$--$10^{-4}$ at $z\sim6$; e.g. \citealt{Fan06})\footnote{Note however even a modest residual fraction can impact the opacity at the systemic redshift (e.g. \citealt{DWH07, LSR11})}.  However, self-shielded systems [damped \lya\ systems (DLAs) and sub-DLAs] inside these HII regions could retain enough neutral hydrogen to imprint strong damping wing absorption.
 If the local ionizing radiation is weak, sightlines through the ionized IGM could have a relatively high incidence of DLAs.  \citet{BH13} recently suggested this can have a large impact on the $z\approx7$ \lya\ fraction.

We model the ionized IGM\footnote{Although \cmfast\ does generate density fields, the perturbation theory \citep{ZelDovich70} approach is inaccurate on the non-linear scales corresponding to DLAs.}
 surrounding \lya\ emitting galaxies with the cosmological hydrodynamic code \enzo\footnote{http://enzo-project.org/} \citep{Enzo13}.
 Our simulation boxes are 22 Mpc on a side.  The root grid is 256$^3$, and we have four additional levels of hydro-refinement resulting in a final baryon resolution of 0.66 proper kpc (pkpc) at $z=7$, which resolves the Jeans length of the relevant systems by a factor of $\sim10$.\footnote{The Jeans length can be written as $L_{\rm J} =$  $7.4 ~ {\rm pkpc} \left( \frac{T}{10^4 {\rm K}} \right)^{0.44} \left( \frac{\Gamma}{0.1} \right)^{-1/3} \left( \frac{\Delta}{10 \Dss} \right)^{-1/2} $, where $\Dss$ is the density at which the gas begins to self-shield, and $\Delta \sim 10 \times \Dss$ is the density of the relevant high-column density systems (see below).}

  In order to allow the gas to dynamically relax, we turn on an optically-thin \citet{HM12} background at $z=9$, roughly corresponding to the midpoint of reionization \citep{Hinshaw13}.  We use the temperature, density, velocity and halo fields at $z=7$, mapped onto a fixed 1024$^3$ grid.  Below we describe our prescription for generating the corresponding neutral hydrogen maps.

As mentioned above, we take the mean photo-ionization rate inside HII regions, $\aveGammaHII$, as our free parameter in computing $\tauHII$. As the clustering of local sources can be important, we construct the local photo-ionization field overdensity, $\Delta_\Gamma({\bf x}) \equiv \Gamma({\bf x})/\aveGammaHII$, where the local ionization rate, $\Gamma({\bf x})$, is computed using a simple, optically thin $r^{-2}$ attenuation profile, and assuming that the galaxy's emissivity of ionizing photons is proportional to its total halo mass (e.g. \citealt{MD08}):
\begin{equation}
\label{eq:sum}
\Gamma({\bf x}) \propto \sum_{i} \frac{M_{\rm halo, i}}{|{\bf x} - {\bf x_i}|^2} ~ e^{-|{\bf x} - {\bf x_i}|/\lmfp} ~.
\end{equation}
Note that the proportionality constant is set by $\aveGammaHII$.  Here ${\bf x_i}$ is the location of halo $i$, $M_{\rm halo, i}$ its mass, and $\lmfp$ the mean free path of ionizing photons.  We take $\lmfp=14$ Mpc, corresponding to the radius of our entire simulation box, which is also in agreement with typical sizes of HII regions near the end of reionization (e.g. \citealt{SM14}).  For computational efficiency, the $\Delta_\Gamma$ field is computed on a lower resolution, 256$^3$ grid.  We stress that this is a conservative\footnote{Throughout this work, we use 'conservative' to indicate assumptions which weaken the derived constraints on reionization.} choice, likely {\it overestimating the importance of DLAs}, as the flux enhancements very near galaxies (where DLAs reside) are smoothed-over using such a relatively coarse grid.

\subsubsection{Self-Shielding Prescription}
\label{sec:SS}

\begin{figure*}
\vspace{-1\baselineskip}
{
\includegraphics[width=0.45\textwidth]{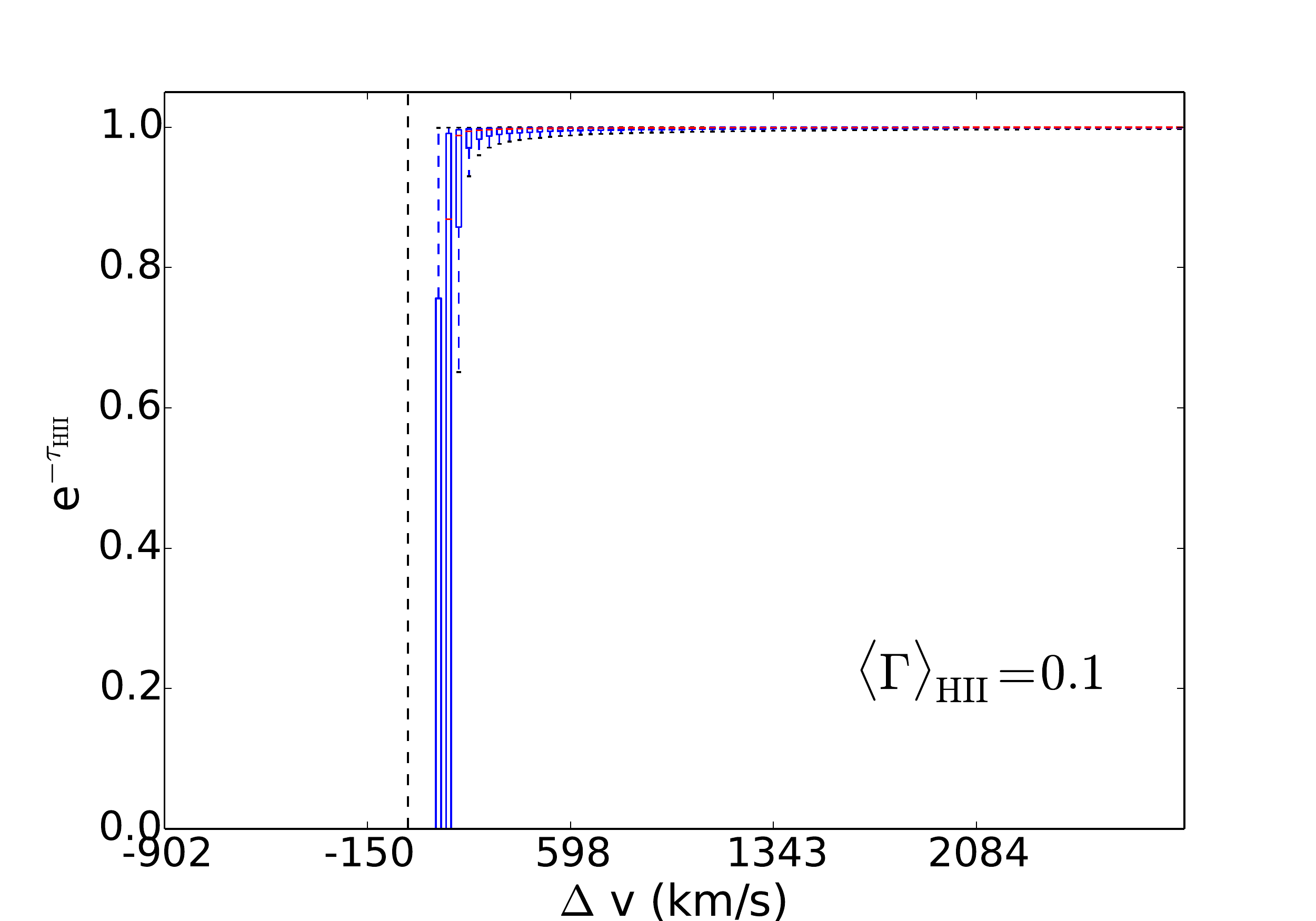}
\includegraphics[width=0.45\textwidth]{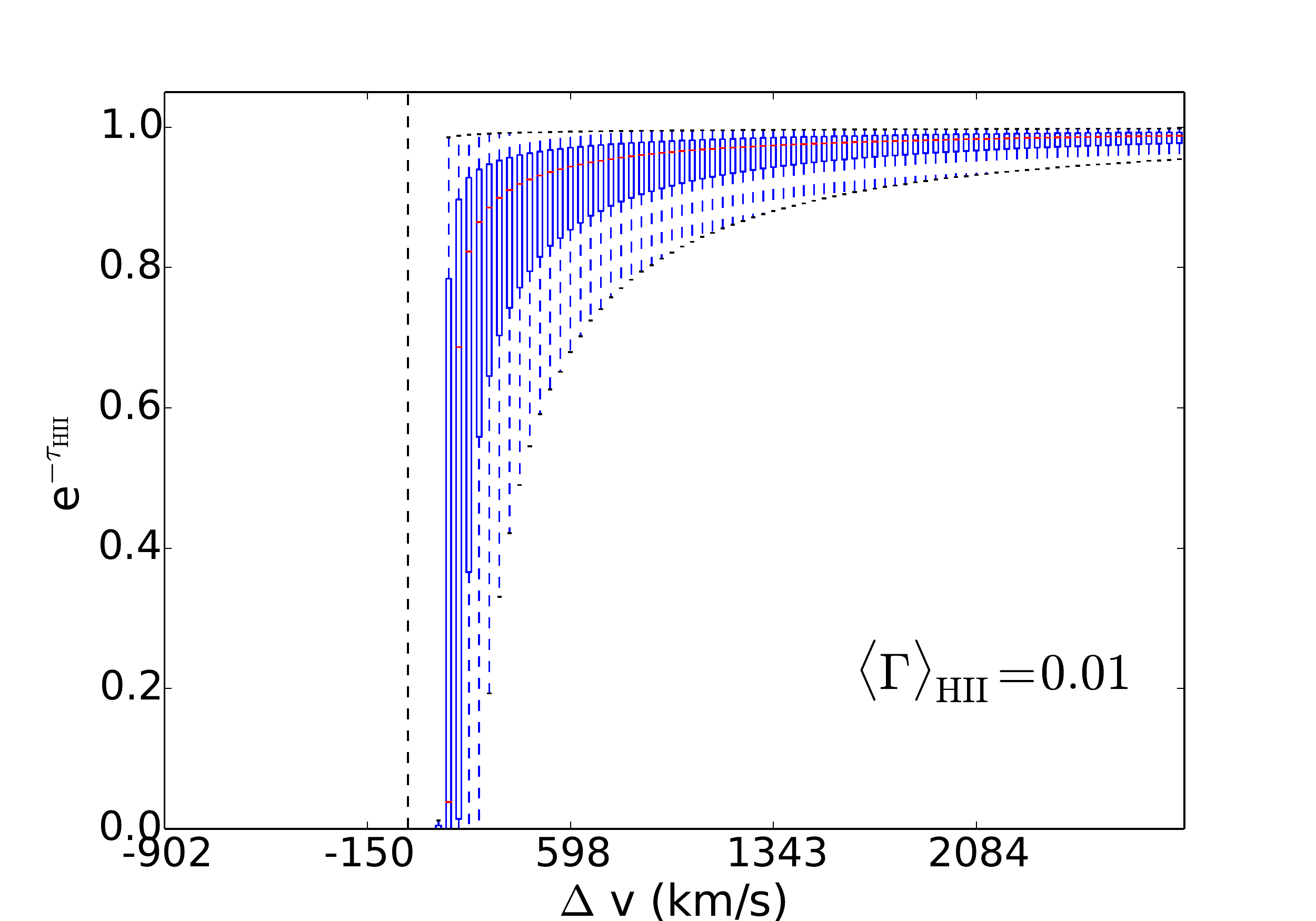}
}
\caption{
Distributions of the damping wing opacity from the ionized IGM ($\QHII=1$), with $\aveGammaHII$=0.1 (0.01) shown on the left (right).
  Red lines correspond to the mean profile, while boxes/whiskers enclose the first two quartiles of the distribution.
}
\label{fig:taud_HII}
\vspace{-0.5\baselineskip}
\end{figure*}

\begin{figure*}
\vspace{-1\baselineskip}
{
\includegraphics[width=0.33\textwidth]{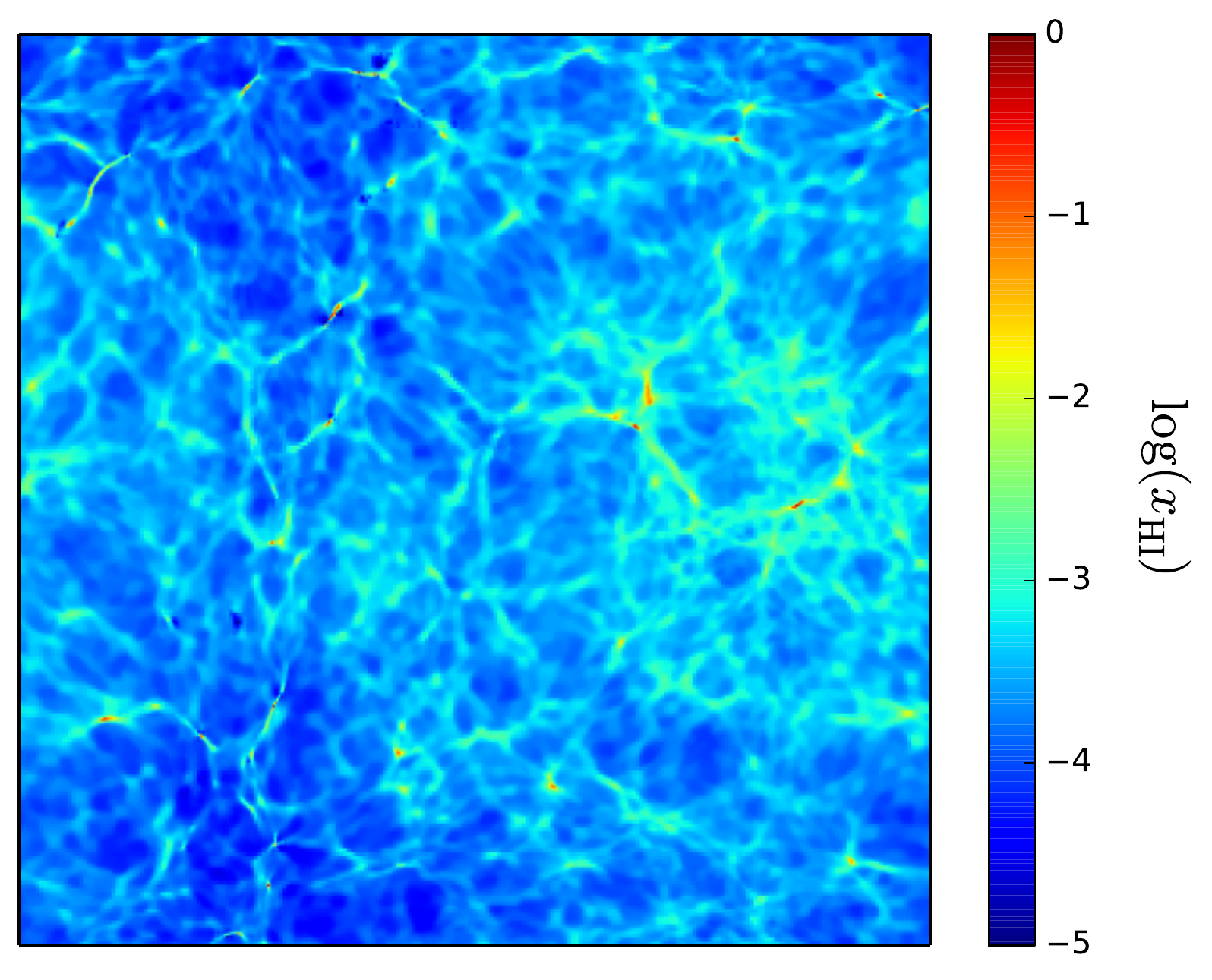}
\includegraphics[width=0.33\textwidth]{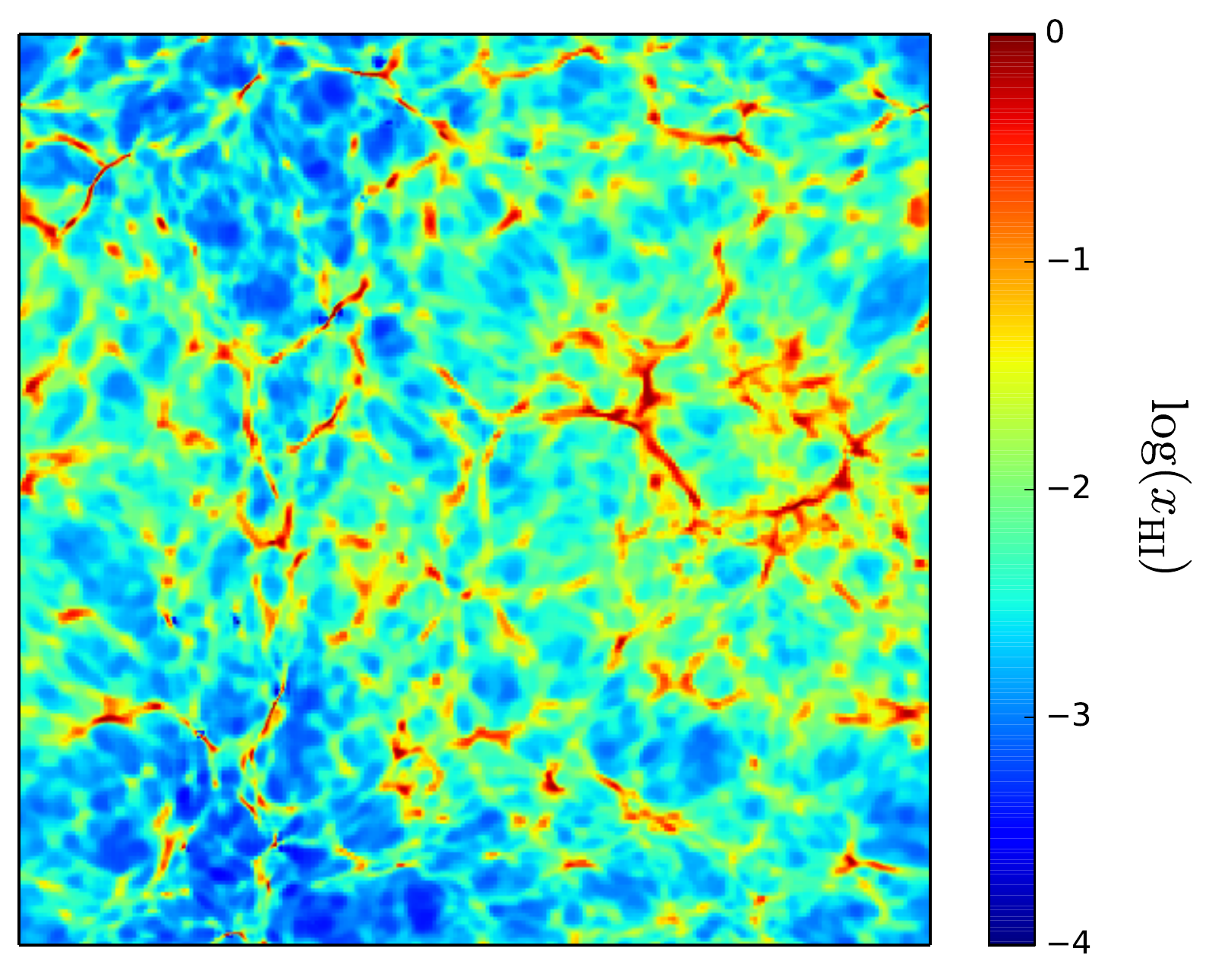}
\includegraphics[width=0.33\textwidth]{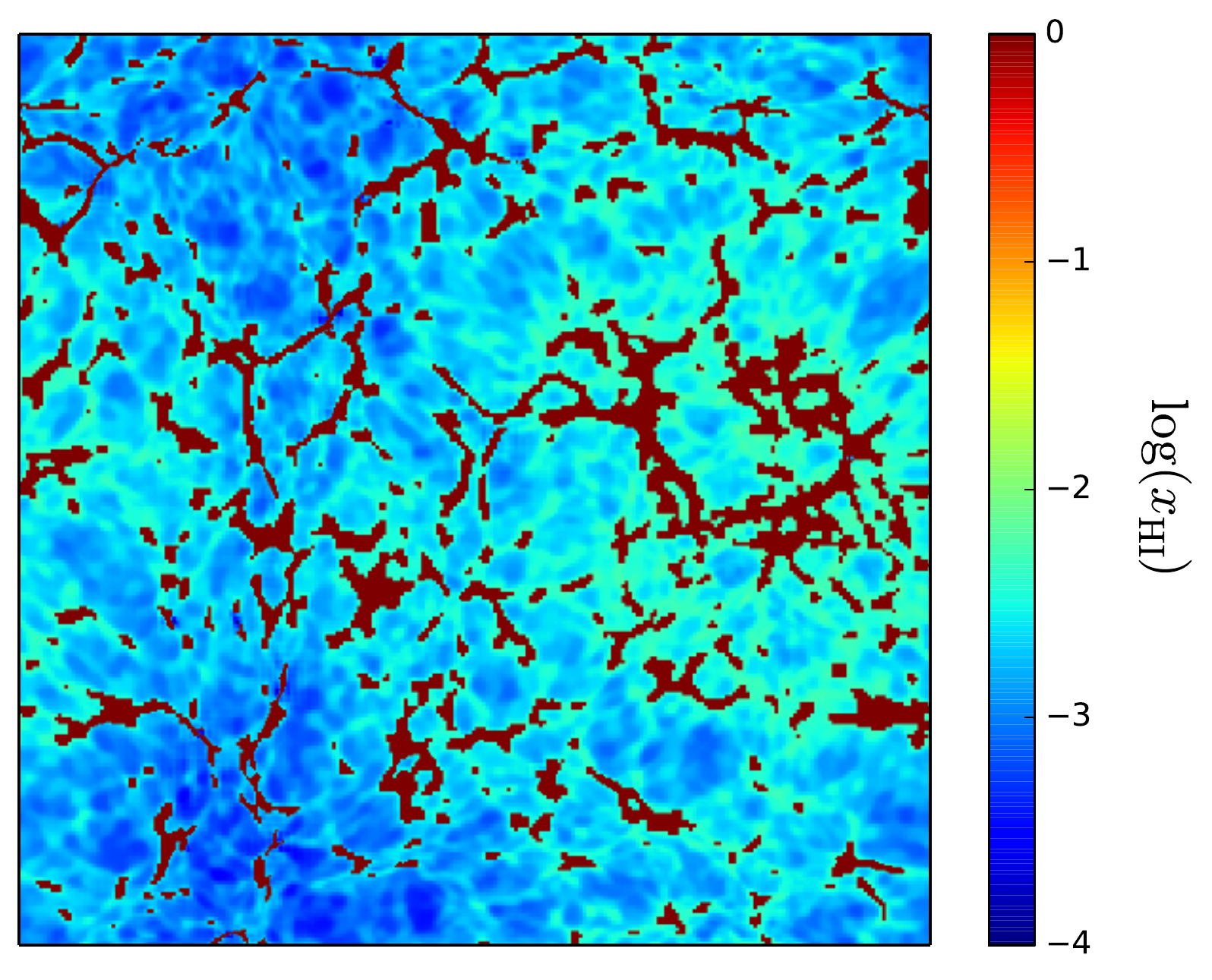}
}
\caption{
Maps of the neutral fraction (22 Mpc across and 21 kpc thick). The first two panels are generated using our fiducial self-shielding prescription, described in the text, and assuming $\aveGammaHII=$ 0.1 ({\it left panel}) and 0.01 ({\it middle panel}).  Like the middle panel, the right panel also corresponds to $\aveGammaHII=0.01$, but instead is computed using the approximation that systems with $\Delta > \Dss$ are fully neutral (c.f. bottom-right panel of Fig. 1 in \citealt{BH13}).
}
\label{fig:StepSS}
\vspace{-0.5\baselineskip}
\end{figure*}

Assuming photoionization equilibrium, we calculate the neutral fraction at a given density:
\begin{equation}
\label{eq:nf}
\nf \Gamma_{\rm ss} = \chi_{\rm HeII} ~ n_{\rm H} ~ (1-\nf)^2 ~  \alpha_{\rm B}(T) ~ ,
\end{equation}
where $n_{\rm H} = \Delta \bar{n}_{\rm H}$ is the hydrogen number density, $\alpha_{\rm B}(T)$ is the case B recombination coefficient (e.g. \citealt{Spitzer78}) for gas at temperature $T$, and $\chi_{\rm HeII} = 1.08$ accounts for singly-ionized helium.  We take into account the self-shielding of the gas through a density-dependent photoionization rate, obtained by an empirical fit to radiative transfer simulations \citep{Rahmati13}:
\begin{align}
\label{eq:UVB}
\frac{\Gamma_{\rm ss}}{\Gamma} & =0.98\times\left[1+\left(\frac{\Delta}{\Delta_{\rm ss}}\right)^{1.64}\right]^{-2.28}+ \nonumber \\
& +0.02\times\left[1+\frac{\Delta}{\Delta_{\rm ss}}\right]^{-0.84}
\end{align}
where $\Delta_{\rm ss}$ is the overdensity above which the gas begins to self-shield \citep{Schaye01}:
\begin{equation}
\label{eq:Dss}
\Dss \approx 15 ~ \left( \frac{\GammaHII}{0.1} \right)^{2/3} \left( \frac{T}{10^4 \text{K}} \right)^{-0.13} \left(\frac{1+z}{7}\right)^{-3} ~ .
\end{equation}
where the pre-factor is computed also assuming a soft, stellar-dominated UV background with an energy index of $\alpha=5$.

In Fig. \ref{fig:taud_HII} we show the resulting distributions of opacities from the ionized IGM ($\QHII=1$).  Comparing these opacities to the  analogous ones sourced by the large-scale reionization morphology, shown in Fig. \ref{fig:taud}, we see that in general the absorption profile from DLAs is steeper than that from the neutral IGM (e.g. \citealt{Miralda-Escude98, MF08damp}).  Moreover, when making reionization constraints it is common to assume that all flux redward of the systemic redshift is fully transmitted through the ionized IGM.  However in Fig. \ref{fig:taud_HII} we see that there is strong absorption from proximate infalling gas at $\Delta v \lsim$100--200 km s$^{-1}$, even if $\QHII=1$ and $\aveGammaHII$ is high (as one might expect at $z\sim6$; see also e.g. \citealt{DMW11}).  One must account for this absorption (present post-reionization) when computing the {\it relative} difference in \lya\ transmission between $z\sim6$ and $z\sim7$ samples.

It is very important to note that gas only {\it begins} to self-shield at $\Delta \sim \Dss$, and is still mostly ionized at these densities.  Strong damping wing absorption requires neutral fractions close to unity, i.e. DLAs rather than LLSs.  Radiative transfer simulations \citep{MOF11, Rahmati13} show that gas becomes mostly neutral well beyond this threshold, requiring $\Delta \gsim 10 \times \Dss$.  Therefore, assuming that systems with $\Delta > \Dss$ are fully neutral, as was done in \citet{BH13}, can dramatically over-estimate the opacity of the ionized IGM (see also \citealt{Keating14}).

This is shown explicitly in Fig. \ref{fig:StepSS}.  In the left and center panels, we show slices through the neutral fraction field, computed according to eq. (\ref{eq:nf})--(\ref{eq:Dss}).  The left panel assumes $\aveGammaHII = 0.1$, while the center one assumes $\aveGammaHII = 0.01$.  As we shall see below, the latter is a conservatively-low choice of $\aveGammaHII$, thus maximizing the importance of self-shielded systems.
  In the right panel, we also assume $\aveGammaHII = 0.01$, but instead compute the neutral fraction with the approximation used in \citet{BH13}: that gas with $\Delta \geq \Dss$ is {\it fully} neutral.  We see that this approximation results in a dramatic discontinuity in the neutral fraction map at the edges of filaments (c.f. bottom right panel of Fig. 1 in \citealt{BH13}; note that the discontinuity is smoothed over by averaging over a fixed step size when creating their Fig. 3).

\begin{figure}
\vspace{-1\baselineskip}
{
\includegraphics[width=0.5\textwidth]{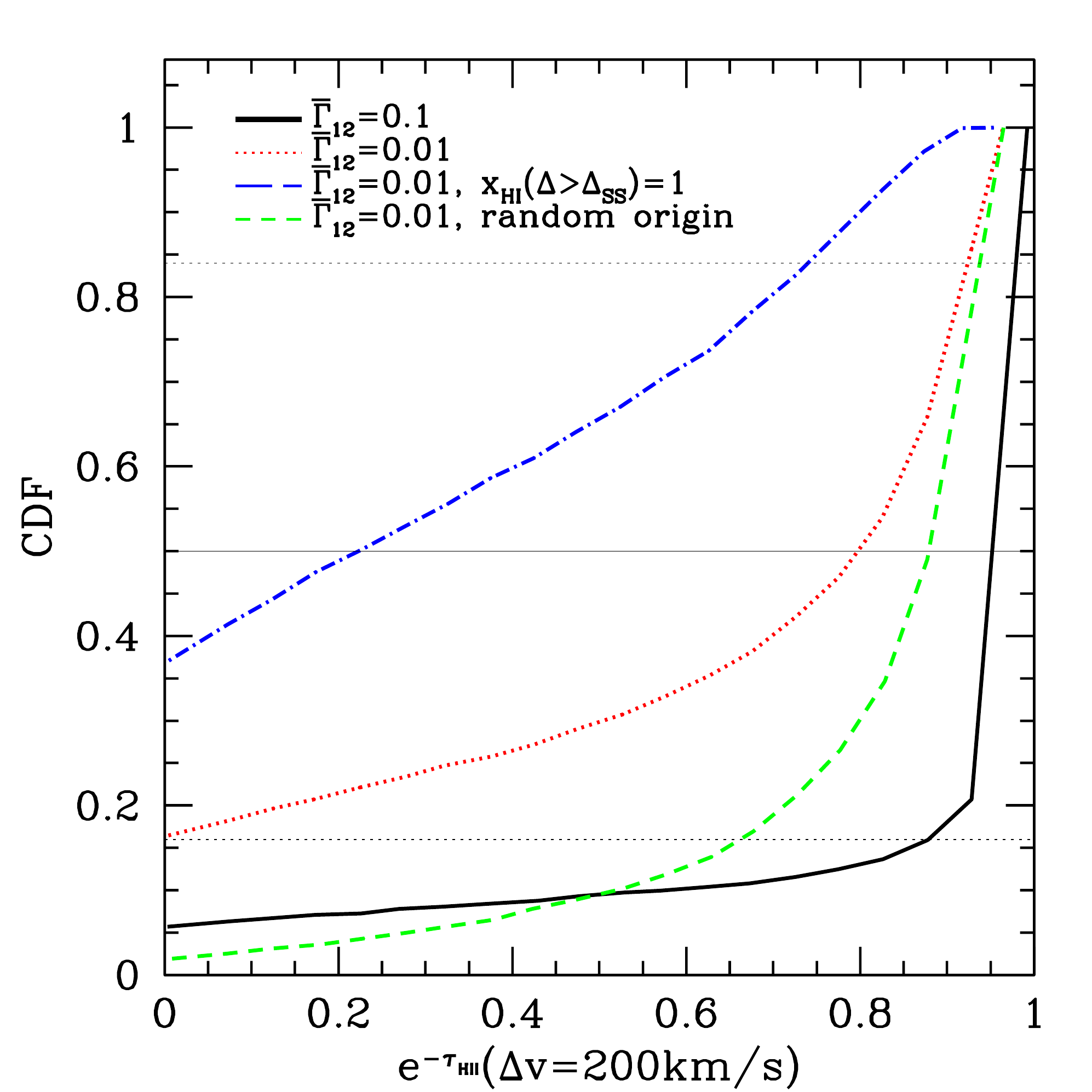}
}
\caption{
Fraction of sightlines having HII opacities less than $\exp[-\tauHII]$, evaluated at $\Delta v$ = 200 km s$^{-1}$ redward of the galaxy's systemic redshift.
  The solid black curve corresponds to $\aveGammaHII$=0.1, while the others correspond to $\aveGammaHII$=0.01.  The dotted-red curve is computed using our fiducial prescription (equations \ref{eq:nf}--\ref{eq:Dss}), while the blue dot dashed curve is computed using the approximation that systems with $\Delta > \Dss$ are fully neutral.  The green-dashed curve is also computed using our fiducial prescription, but with sightlines originating at random locations in the IGM, instead of halo centers.
}
\label{fig:HII_CDFs}
\vspace{-0.5\baselineskip}
\end{figure}

This is further quantified in Fig. \ref{fig:HII_CDFs}, where we show the CDFs of the transmission, $\exp[-\tauHII]$, again computed at a rest-frame velocity offset of $\Delta v=$ 200 km s$^{-1}$.
  The red dotted curve is computed according to our fiducial prescription above assuming $\aveGammaHII=0.01$, corresponding to the middle panel of Fig. \ref{fig:StepSS}.  Even for such a low $\aveGammaHII$, the attenuation along most sightlines is modest, with the average transmission at $\Delta v=$ 200 km s$^{-1}$ being $\exp[\tauHII]\approx0.8$.  There is a high opacity tail corresponding to $\sim$15\% of LOSs which encounter high-column density systems.  On the other hand, when the opacity is computed with the previous simplification of fully neutral gas at $\Delta \geq \Dss$ ({\it blue dot-dashed curve}), the transmission drops dramatically.  In this case, the mean transmission at $\Delta v=$ 200 km s$^{-1}$ is $\exp[\tauHII]\approx0.2$, and the CDF is very broad.

It is worthwhile also to highlight the qualitative differences between the CDFs in figures \ref{fig:reion_CDFs} and \ref{fig:HII_CDFs}.  Namely, the opacities sourced from the HII regions (i.e. self-shielded systems) are relatively bimodal: either a LOS encounters a nearby high-column density system and the \lya\ is strongly absorbed, or it does not and the attenuation is modest.  On the other hand, opacities sourced from the large-scale reionization morphology are unimodal, as the damping wings from the cosmic neutral patches are smoother functions of wavelength and so even distant patches impart a modest attenuation. This is in contradiction to recently-popularized Bayesian studies (e.g. \citealt{Treu12, Treu13, Pentericci14, Tilvi14}) which use an empirical 'on/off' model (called 'patchy') as a proxy for reionization attenuation.


\subsubsection{Biases in the density and ionization fields}
\label{sec:biases}

Galaxies reside in biased locations of both the density and ionizing background fields.  In terms of the resulting IGM opacity, these two biases act in opposite directions.  Biased locations of the density field imply more surrounding structure, capable of hosting high-column density systems.  On the other hand, biased locations in the photoionization field imply a stronger overdensity criterion for self-shielding (i.e. higher value of $\Dss$ in eq. \ref{eq:Dss} resulting from a higher $\Gamma$).

We briefly show the impact of these biases on the \lya\ transmission by comparing the red-dotted and green-dashed lines in Fig. \ref{fig:HII_CDFs}.  The former is constructed with LOSs originating from halos, while the latter is constructed from LOSs originating from random locations in the simulation box.  We see that in this case, the density bias ``wins'': although the mean transmissions are comparable, sightlines originating at random locations are understandably far less likely to encounter nearby DLAs.  The increase in the local photo-ionization rate is unable to counter the higher incidence of structures near galaxies.  However, we again caution that we underestimate the fluctuations in the local photo-ionization rate, since (i) the photo-ionization overdensity field, $\Delta_\Gamma$, is computed on a relatively coarse, 256$^3$ grid; and (ii) our \enzo\ box is too small to capture large-scale fluctuations in $\Gamma$ (e.g. \citealt{Crociani11}). 

\subsection{The intrinsic Ly$\alpha$ emission line}
\label{sec:line}

Having constructed a database of optical depth profiles from {\it both} reionization, $\taure$, {\it and} the local HII region, $\tauHII$, we now need the intrinsic \lya\ line, $J(\nu)$, emerging from the galaxy's ISM and CGM (which in our case corresponds to distances within 0.16 cMpc of the galaxy, as mentioned above).  The total transmission, $T_{\rm IGM}$, is then an integral over the intrinsic line:
\begin{equation}
\label{eq:T}
T_{\rm IGM} = \int d\nu J(\nu) \exp[-\taure(\nu) - \tauHII(\nu)] ~ ,
\end{equation}
\noindent where $J(\nu)$ is normalized to integrate to unity.

Modeling $J(\nu)$ is beyond the scope of this work.  Hence we just assume simple Gaussian profiles, centered at $\Delta v = 0$, 200, 400 km s$^{-1}$ redward of the systemic redshift,
with a r.m.s. width of 100 km s$^{-1}$, roughly corresponding to the circular velocities of the host halos (e.g. \citealt{BL04}).  In reality, the profile of $J(\nu)$ is much more complicated, likely involving radiative transfer through outflowing material (see for example \S 5 in \citealt{DMW11} and references therein).  However, our relatively-narrow Gaussians sample the range of \lya\ emission observed in low-$z$ LAEs (e.g. \citealt{Steidel10, Shibuya14}).  By sampling both the low and high ends of the likely systemic offset of the \lya\ profile, we bracket the expected impact of $J(\nu)$ on our results.
  As we shall see below, the choice of $J(\nu)$ does not have a large impact on $Q_{\rm HII}$ constraints (unless there is {\it evolution} in $J(\nu)$ from $z=6\rightarrow7$)\footnote{The line profile might indeed evolve with redshift.  For example, \citet{JSE12} show that the covering factor of low-ionization absorbers (which trace HI) decreases from $z=3\rightarrow4$ in LBGs. They argue that it is in fact the covering factor that is regulating \lya\ escape, which could imply a redshift evolution of $J(\nu)$.}
 though it does have a modest impact on $\aveGammaHII$ constraints.

\section{Results}
\label{sec:results}

\subsection{Total IGM transmission}
\label{sec:tot}

\begin{figure*}
\vspace{-1\baselineskip}
{
\includegraphics[width=0.45\textwidth]{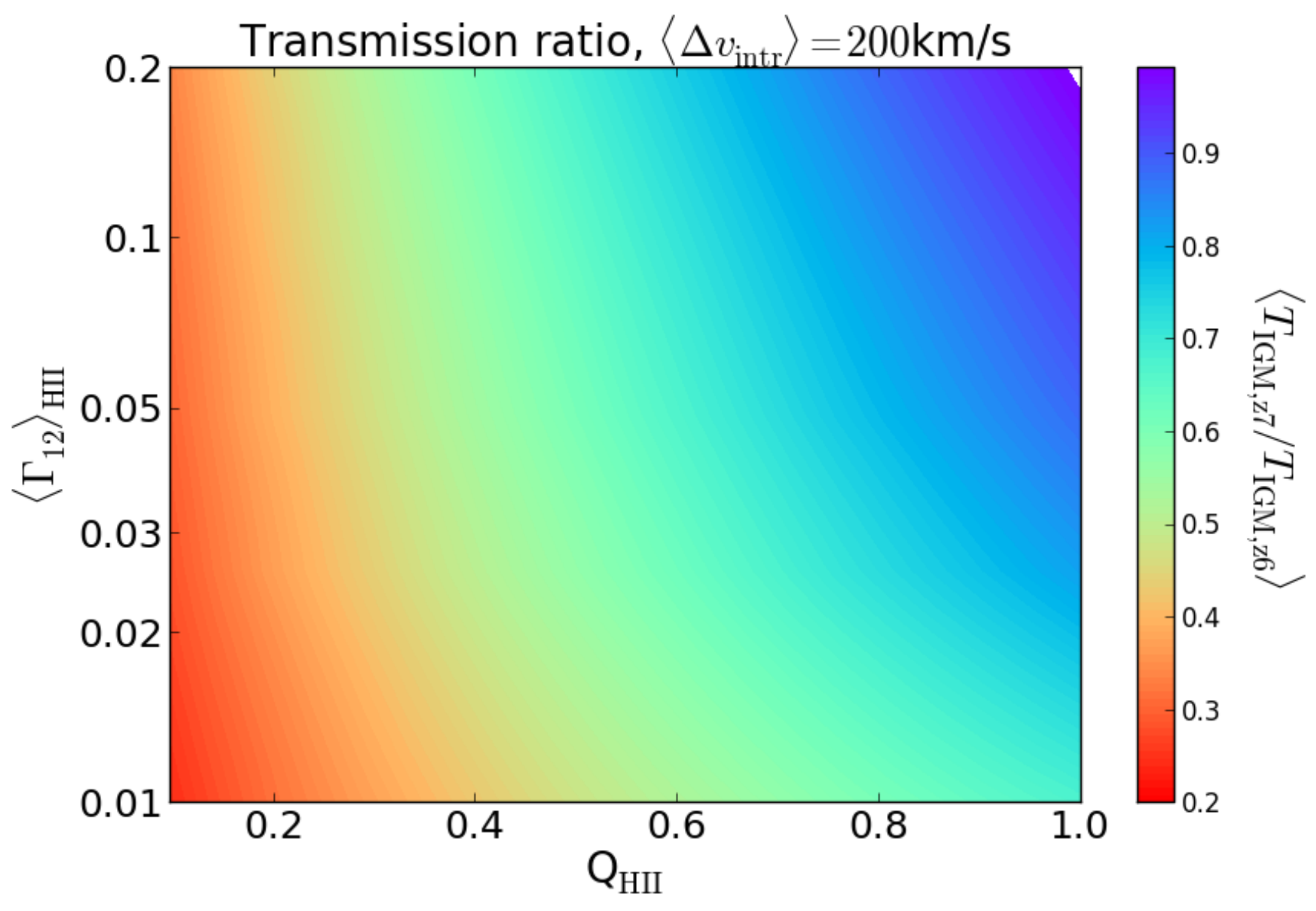}
\includegraphics[width=0.45\textwidth]{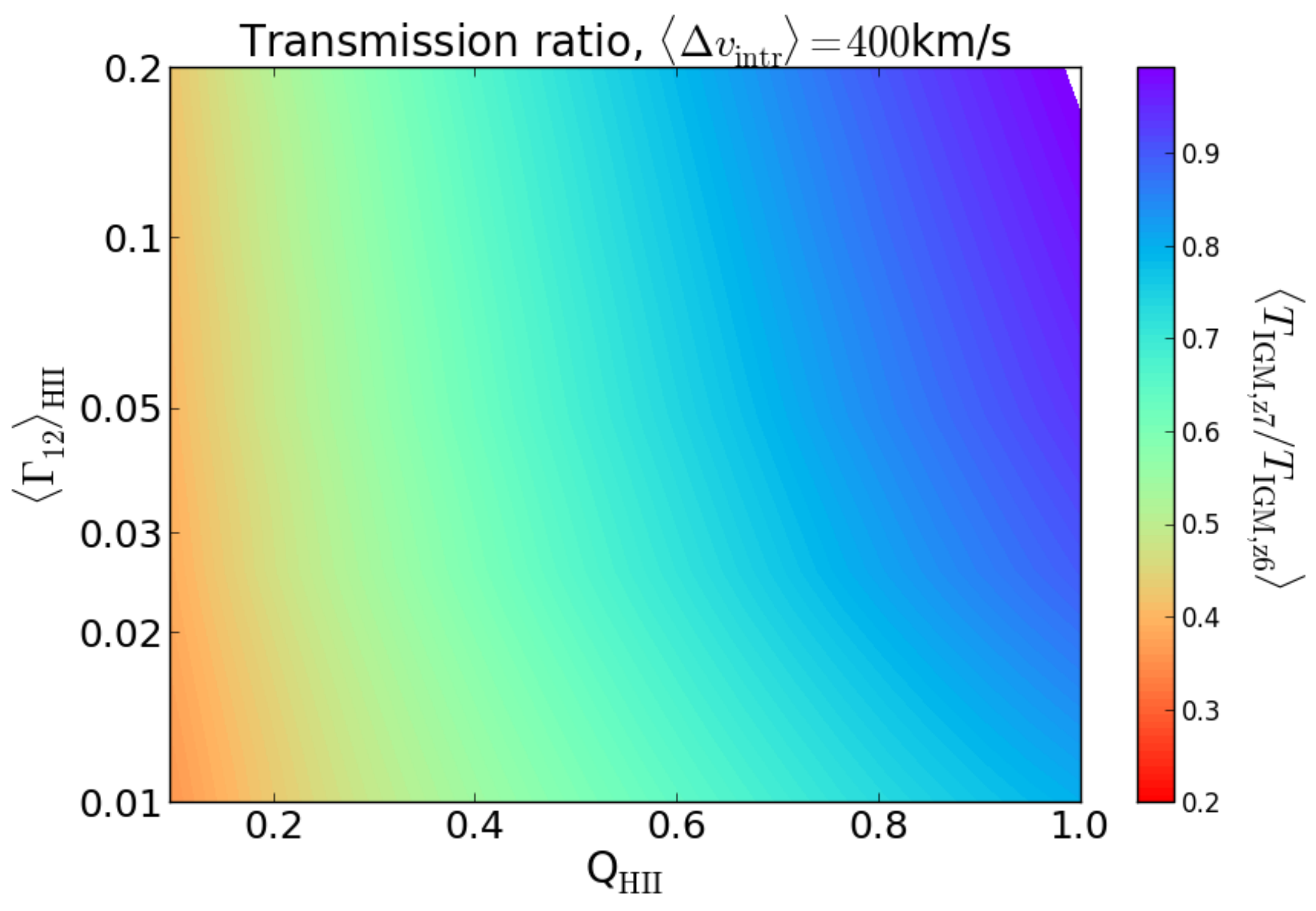}
}
\caption{
Sightline-averaged ratio of IGM \lya\ transmission (eq. \ref{eq:T}) at $z=6$ and $z=7$:  $\langle T_{\rm IGM, z7}/T_{\rm IGM, z6} \rangle$.  The axes correspond to our $z=7$ model parameters: $(\QHII, \aveGammaHII)_{z7}$.  $T_{\rm IGM, z6}$ is computed assuming $(\QHII, \aveGammaHII)_{z6} = (1, 0.2)$, and neglecting redshift evolution of other quantities (i.e. by construction $T_{\rm IGM, z7}/T_{\rm IGM, z6} = 1$ in the top right corner of the panels).  
The left (right) panel assumes an intrinsic emission profile, $J(\nu)$, centered at $\Delta v =$ 200 (400) km s$^{-1}$.
}
\label{fig:Tratio}
\vspace{-0.5\baselineskip}
\end{figure*}

 In Fig. \ref{fig:Tratio} we plot the sightline-averaged ratio of IGM \lya\ transmission (eq. \ref{eq:T}) at $z=7$ and $z=6$:  $T_{\rm IGM, z7}/T_{\rm IGM, z6}$.  $T_{\rm IGM, z6}$ is computed assuming $(\QHII, \aveGammaHII)_{z6} = (1, 0.2)$, and neglecting redshift evolution of other quantities (i.e. $\langle T_{\rm IGM, z7}/T_{\rm IGM, z6} \rangle \equiv 1$ in the top right corner of the parameter space; see below for some motivation of this conservative choice).  The left (right) panel assumes an intrinsic emission profile, $J(\nu)$, centered at $\Delta v =$ 200 (400) km s$^{-1}$.

From the modest inclination of the isocontours over the upper half of parameter space in Fig. \ref{fig:Tratio}, we see that the transmission ratio is more sensitive to $\QHII$ than $\aveGammaHII$, despite our conservative assumptions mentioned above.  Proximate self-shielded systems only impact the average transmission when $\aveGammaHII \lsim 0.02$.  The transmission is even less sensitive to $\aveGammaHII$ if the intrinsic \lya\ line, $J(\nu)$, has a larger systemic velocity offset (i.e. the right panel of Fig. \ref{fig:Tratio}).  As already mentioned, this is due to the fact that self-shielded systems generally have absorption profiles which are steeper with wavelength than the neutral IGM (e.g. \citealt{Miralda-Escude98, MF08damp, McQuinn08}).  Therefore if the intrinsic \lya\ emission has a significant contribution far redward of the systemic redshift, it will be even less sensitive to self-shielded systems than reionization.  


It is also interesting to note that even for the largest $z=6\rightarrow7$ evolution considered, $(\QHII, \aveGammaHII)_{z7} \approx (0.1, 0.01)$\footnote{A fully neutral (or close to fully neutral) universe would result in lower transmission.  We do not however explore such tiny values of $\QHII$ since (i) our approach does not model tiny, sub-grid HII regions prevalent in the very first stages of the EoR; (ii) the implied sharp reionization ($\Delta z_{\rm re} \approx 1$) is extremely unlikely.}, the transmission ratio is still not very small, $\langle T_{\rm IGM, z7}/T_{\rm IGM, z6} \rangle =$ 0.2--0.4.  This is driven by the emission redward of the systemic velocity (physically motivated by radiative transfer through outflows; e.g. \citealt{Verhamme08}), some of which is transmitted even through a neutral IGM.

We stress that up to now, our results have not made use of any observations.  Figure \ref{fig:Tratio} shows our model predictions, {\it which can be used to interpret evolving observational data sets, as we illustrate below.}

\subsection{Lyman alpha fractions}
\label{sec:lya_frac}


We now turn to modeling the redshift evolution of the \lya\ fraction.  For this, we need the distribution of rest frame equivalent widths (REWs) at $z\approx6$, REW$_{z6}$.  Our $z\approx6$ galaxy sample consists of 56 color-selected galaxies with VLT spectroscopy, and is a collection of past and ongoing programs (\citealt{Fontana10, Pentericci11, Pentericci14}, Vanzella et al. in preparation). The integration times span the range between 5 to 18 hours on target.  The sample does not include galaxies for which there is no spectroscopic redshift confirmation (either from a \lya\ line or from the \lya\ break).  Since such null detections are included when presenting $z\approx7$ fractions, we expect our $z\approx6$ spectroscopic sample to over-predict the $z=6\rightarrow7$ drop.  Hence, we add four additional REW=0\AA\ galaxies, so that our fiducial $z\approx6$ fractions match those recently published in \citet{Schenker14}, allowing us to compare against their $z\approx7$ results.  We note that our main predictions of the $z=6$ and 7 \lya\ fraction ratios in Fig. \ref{fig:ratio_faint} are unaffected by this overall normalization, {\it depending only on the REW$_{z6}>25$ \AA\ distribution of UV faint galaxies}.  

\begin{figure*}
\vspace{-1\baselineskip}
{
\includegraphics[width=0.33\textwidth]{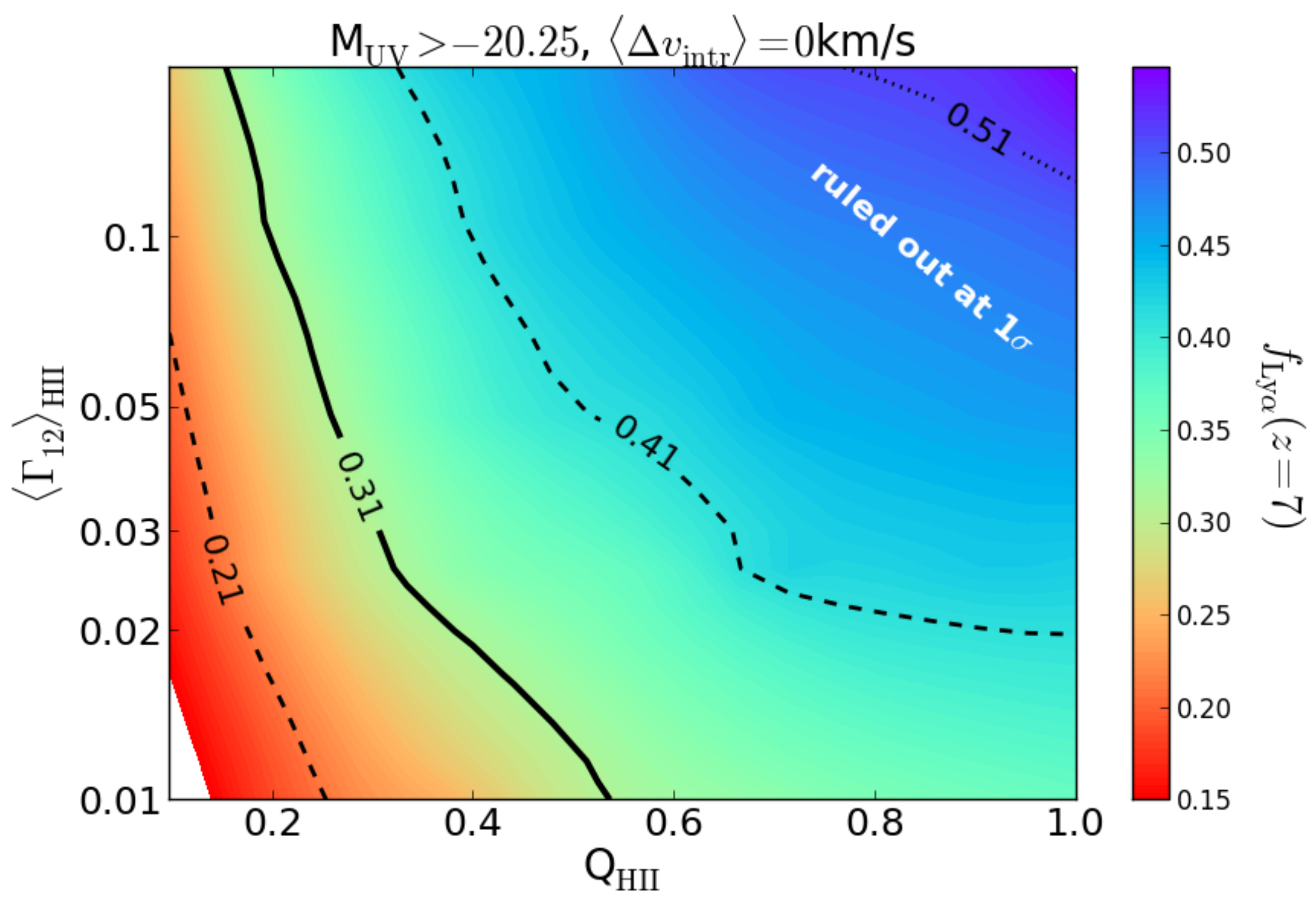}
\includegraphics[width=0.33\textwidth]{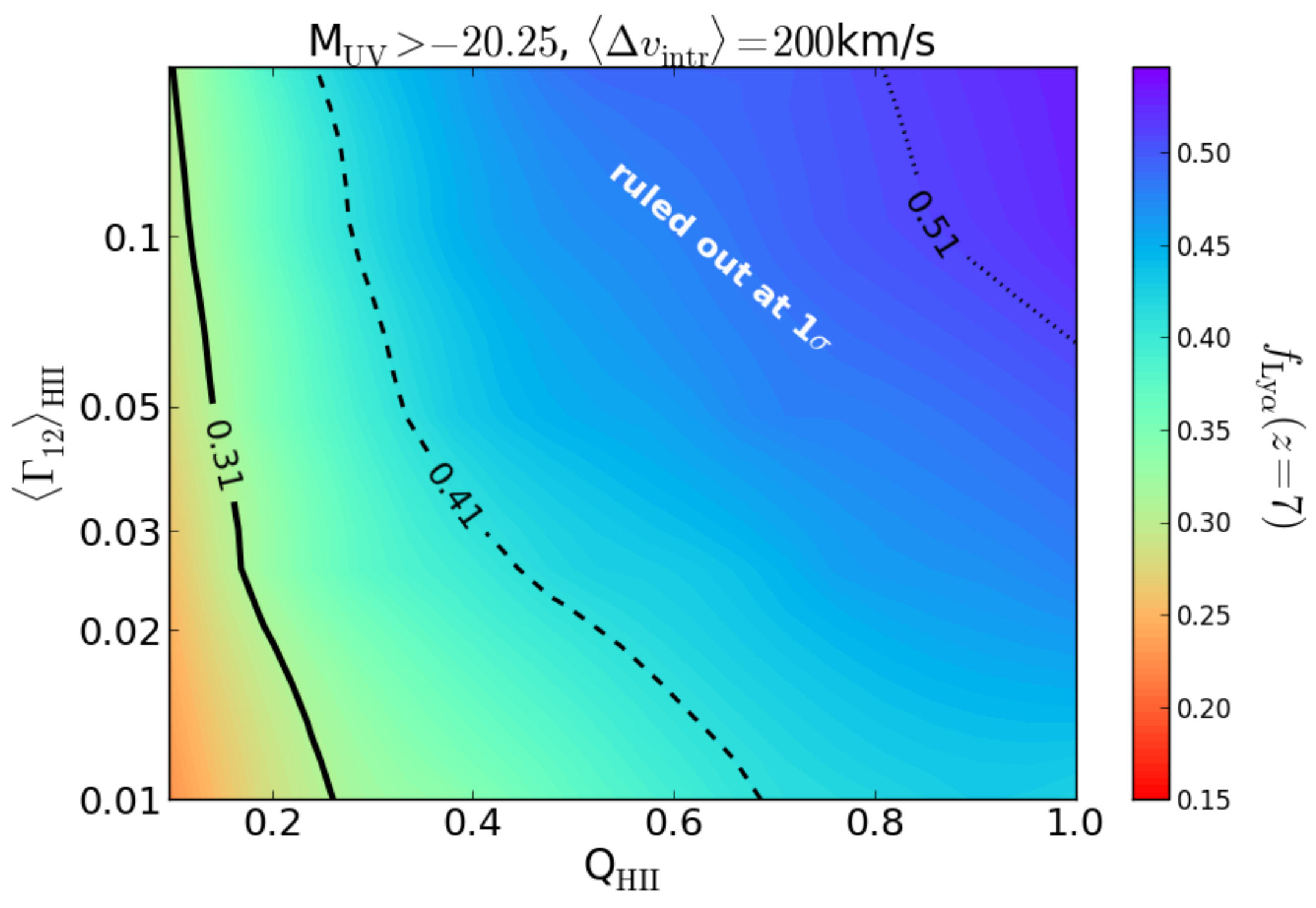}
\includegraphics[width=0.33\textwidth]{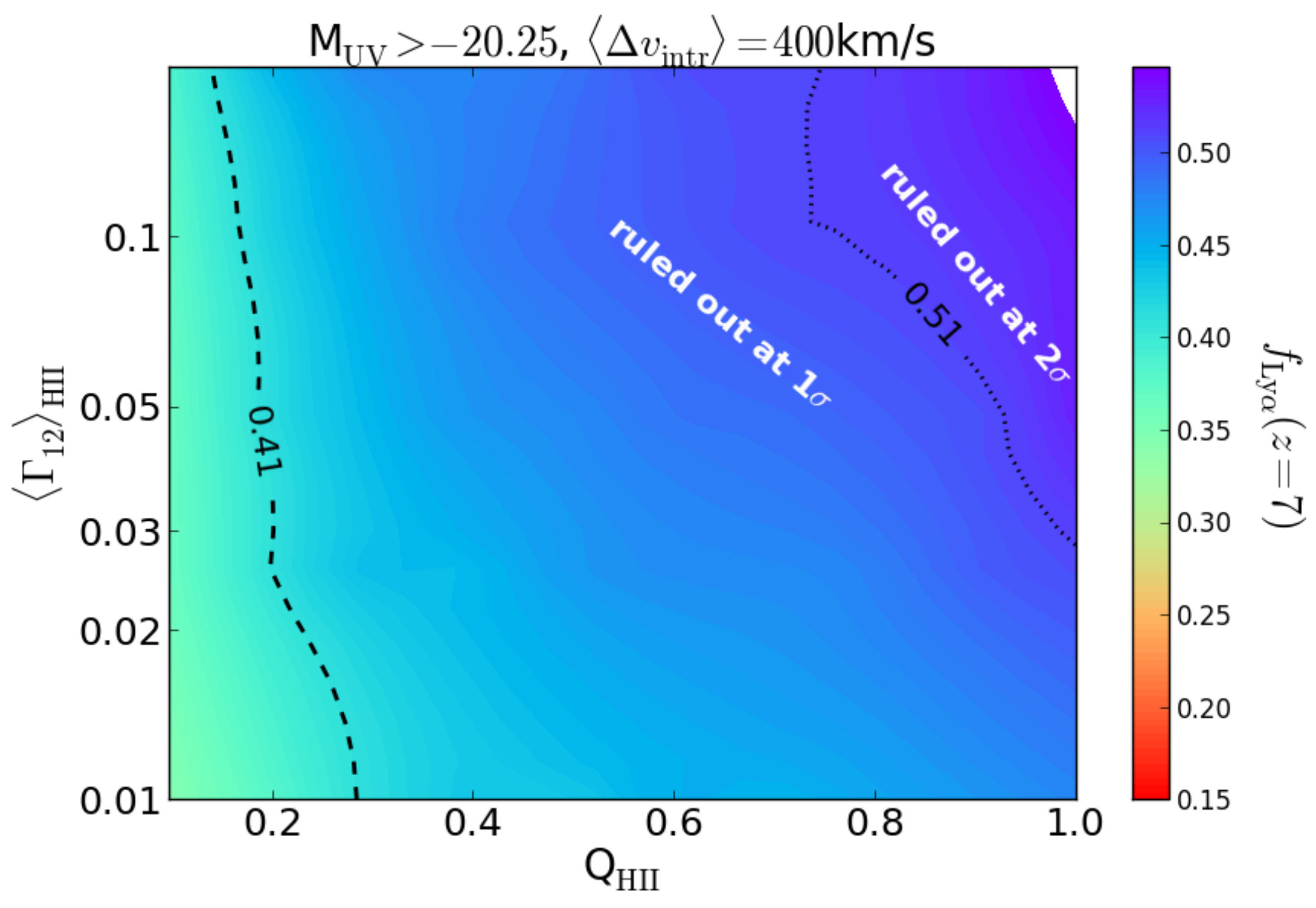}
}
\caption{
Fraction of UV faint galaxies (M$_{\rm UV}\gsim -20.25$) with REW$>25$\AA\ at $z=7$, assuming the intrinsic REW distribution is the same as at $z=6$.
We denote the observed estimates of $f_{\rm Ly\alpha}(z=7)$ from \citet{Schenker14} ({\it thick solid lines}), as well as the 1 (2) $\sigma$ iso-contours with thin dashed (dotted) lines.
The left, center, right panels assume an intrinsic emission profile, $J(\nu)$, centered at $\Delta v =$ 0, 200, 400 km s$^{-1}$, respectively.
}
\label{fig:Lya_frac}
\vspace{-0.5\baselineskip}
\end{figure*}

In Fig. \ref{fig:Lya_frac}, we plot the $z\approx7$ \lya\ fraction, $f_{\rm Ly\alpha}(z\approx7)$, defined as the fraction of UV faint\footnote{As already noted (e.g. \citealt{Stark10, Ono12}), the most dramatic change in the \lya\ fraction from $z=6\rightarrow7$ is driven by UV faint galaxies.} (M$_{\rm UV}\gsim -20.25$) galaxies with REWs greater than 25 \AA.    This is computed by Monte Carlo sampling the rest frame equivalent width compilation of $z\approx6$ galaxies, REW$_{z6}$
  Each sightline, $i$, at $z=7$ is then assigned a REW$^i_{z7}$=REW$^i_{z6}$$\times T^i_{\rm IGM, z7}/T^i_{\rm IGM, z6}$, and we compute the resulting fraction with REW$_{z7}>25$\AA\ for each point in our $(\QHII, \aveGammaHII)_{z7}$ parameter space.

 We denote the observed estimates of $f_{\rm Ly\alpha}(z=7)$ from \citet{Schenker14} ({\it thick solid lines}), as well as the 1 (2) $\sigma$ iso-contours with thin dashed (dotted) lines.
 We stress that this analysis assumes that the change in IGM transmission is entirely driving the change in \lya\ fraction.
 For simplicity, we do not correlate the REW with the host halo mass.  This is reasonable, as our host halos only span a factor of few in mass and the halo bias (i.e. local environment) is relatively constant over this mass range.  Moreover, we neglect cosmic variance in the reionization morphology, which \citet{TL13} showed can impact the interpretation of previous observational samples.  This choice was made for several reasons: (i) new samples (e.g. \citealt{Pentericci14, Schenker14}) probe much larger volumes, better sampling reionization morphology (especially during the relevant advanced stages); (ii) calculations for the newer samples in \citet{Pentericci14} already showed EoR cosmic variance to be small, especially when computed redward of the systemic redshift, as relevant for most of the intrinsic \lya\ line profile; (iii) we wish to present general results which can be used to easily interpret new data.


If the intrinsic \lya\ line has a large systemic redshift, $\Delta v =$ 400 km s$^{-1}$ ({\it right panel}), it is difficult for the IGM to significantly impact the transmission, as most of the emission lies far out on the damping wing tail of the absorption.  Hence, a very large neutral fraction is required for even a modest change in the \lya\ fraction.

  On the other hand, under the more reasonable assumption that the intrinsic line is less redshifted, $\Delta v =$ 200 km s$^{-1}$ ({\it center panel}), we see that $\QHII(z=7)\lsim0.25$\footnote{For simplicity, it is common in the literature to assume $T_{\rm IGM, z6}$=1 redward of the systemic redshift. However resonant absorption in infalling material can attenuate intrinsic emission redward of the systemic redshift, even without reionization or DLAs.  Compared to such prior analysis, our more realistic treatment results in higher $T_{\rm IGM, z7}/T_{\rm IGM, z6}$ ratios, strengthening constraints on $\QHII$ (i.e. by making it more difficult for the IGM to produce a given drop in the \lya\ fraction).  This offsets the loosening of constraints from our new, feedback-limited reionization morphologies.   Therefore, our derived constraints on $\QHII$ (assuming no $\aveGammaHII$ evolution), are not as different from previous ones, as would be expected from the discussion in \S \ref{sec:recomb}.  We also note that models in which outflows set the emerging \lya\ line profile result in considerable line shifts, even for modest outflow velocities (e.g. \citealt{DMW11}).  Hence the above distinction is less relevant for these models.}
is required to be consistent with observations at 1$\sigma$, barring evolution in $\aveGammaHII$ (c.f. the top axis of the panel).  Conversely, even $\aveGammaHII(z=7)\sim0.01$ is not low enough to be consistent with observations at 1$\sigma$, if the IGM is ionized at $z=7$ (c.f. the right axis of the panel).  As we shall see below, the IGM properties are highly unlikely to evolve so dramatically over such a narrow redshift interval.

On the other hand, allowing for a {\it joint} evolution in IGM properties relaxes somewhat the mild tension with data.  For example, $\QHII\lsim0.5$ and $\aveGammaHII \lsim 0.02$ are allowed at 1$\sigma$ in the middle panel.

The left panel shows \lya\ fractions under the extreme assumption that there is no offset between the intrinsic \lya\ line and the systemic redshift of the galaxy.  This model is inconsistent with observations at lower redshifts (e.g. \citealt{Steidel10, Shibuya14}).  It results in only a few percent of the \lya\ line being transmitted far out on the red wing of the line, even at $z\approx6$, as infalling gas on average resonantly absorbs the emission at $\Delta v\approx$ 100-200 km s$^{-1}$ redward of the systemic redshift (c.f. the left panel of Fig. \ref{fig:taud_HII}).  However, it serves to illustrate that for the majority of 'reasonable' parameter space (i.e. the upper right quadrant), the \lya\ fractions are only mildly different from the more reasonable $\Delta v =$ 200 km s$^{-1}$ model.  This is due to the fact that we are normalizing the \lya\ fractions to the same value at $(Q_{\rm HII}, \aveGammaHII)$ = $(1, 0.2)$, our fiducial choice for $z\approx6$.



\subsection{Evolution of the Lyman alpha fractions}
\label{sec:lya_ratio}

\begin{figure}
\vspace{-1\baselineskip}
{
\includegraphics[width=0.5\textwidth]{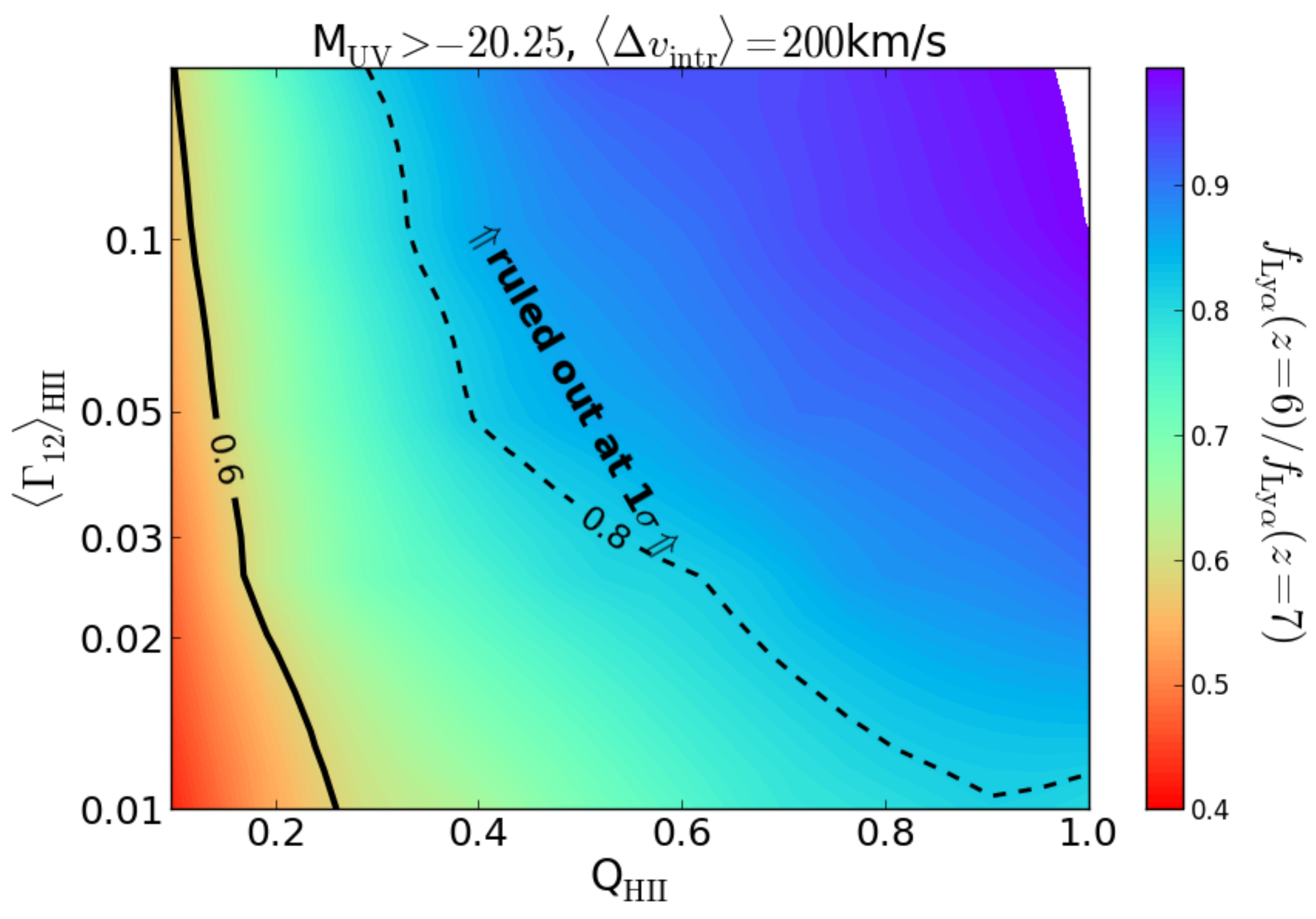}
}
\caption{
Ratio of $z\approx6$ and $z\approx7$ \lya\ fractions, $f_{\rm Ly\alpha}(z=6)/f_{\rm Ly\alpha}(z=7)$, for the UV faint sample, M$_{\rm UV} > -20.25$, assuming $\Delta v =$ 200 km s$^{-1}$. The black solid (dashed) curves corresponds to the best fit (1$\sigma$) from \citet{Schenker14}.
}
\label{fig:ratio_faint}
\vspace{-0.5\baselineskip}
\end{figure}

The {\it evolution} of the \lya\ emission is better understood as a {\it ratio} of the \lya\ fractions.  In Fig. \ref{fig:ratio_faint} we show the ratio of the $z\approx6$ and $z\approx7$ \lya\ fractions for the UV faint galaxies: $f_{\rm Ly\alpha}(z\approx6)/f_{\rm Ly\alpha}(z\approx7)$.  Here for simplicity, we only assume $\Delta v =$ 200 km s$^{-1}$ for the intrinsic \lya\ emission. 

Fig. \ref{fig:ratio_faint} can be readily applied to interpret new observations.  The ratios shown in the figure do not depend on the assumed normalization of the $z\approx6$ REW PDF, nor on how the REW is distributed at low values REW$<25$\AA\ (which is difficult to determine observationally), as these REW are already 'non-detections' by the definition of the \lya\ fraction.  It only depends on the shape of the $z\approx6$ distribution at REW$>25$\AA, as these are the galaxies who can move over from the 'detectable \lya' category to the 'non detectable \lya' category, due to IGM evolution from $z\approx6$ to 7.  

This ratio of \lya\ fractions shows the $z=6\rightarrow7$ evolution more explicitly.  For our parameter space, the \lya\ fraction physically cannot drop by more than a factor of $\sim2$.  This is driven by objects with large values of REW$_{z6}$.
  The evolution of IGM properties (Fig. \ref{fig:Tratio}) is insufficient to suppress such high REWs to values of REW$_{z7}<25$ \AA. 
 As we discuss below, more reasonable models lie in the upper right quadrant of parameter space; here the ratio of \lya\ fractions only evolves by tens of percent.  Larger evolution of the \lya\ fraction would be possible if the distribution of REW$_{z6}$ was more sharply peaked towards low values, REW$_{z6}\approx25$ \AA, again highlighting the need for larger, more robust galaxy samples.

We also show the 1$\sigma$ observational uncertainty on the ratio of \lya\ fractions, computed with standard error propagation, with the dashed curve.  Unlike the observational uncertainty on $f_{\rm Ly\alpha}(z\approx7)$ shown in Fig. \ref{fig:Lya_frac}, the uncertainty in $f_{\rm Ly\alpha}(z\approx6)/f_{\rm Ly\alpha}(z\approx7)$ includes the additional Poisson error in $f_{\rm Ly\alpha}(z\approx6)$.   The black solid (dashed) curves corresponds to the best fit (1$\sigma$) from \citet{Schenker14} (see their Fig. 9). {\it All of the parameter space is consistent with observations at 2$\sigma$.}\footnote{We do not compare with recent null detections at $z\approx8$ (e.g. \citealt{Treu13,  Schenker14}), as the resulting constraints depend heavily on understanding the systematic uncertainties and error bars on the null detection.  For example, taken at face value from Fig. 9 in \citet{Schenker14}, their claimed evolution of $z=6\rightarrow8$ is inconsistent with IGM attenuation by more than 5$\sigma$, while non-Bayesian upper limits at $z\approx8$ (e.g. \citealt{Treu13, Tilvi14}) are fully consistent with no evolution over this redshift range.}


One can also convert Fig. \ref{fig:ratio_faint} to a likelihood and marginalize over one of the dimensions.  Using a standard $e^{-\chi^2/2}$ estimator and adopting a uniform prior in $\log \aveGammaHII$, we obtain a marginalized 68\% confidence level (C.L.) of $Q_{\rm HII}(z\approx7) \leq0.6$. Future measurements could be used to motivate priors in either $\aveGammaHII$ or $Q_{\rm HII}$, further increasing the constraining power of this analysis.


\subsection{What is ``reasonable'' for the IGM at z = 7?}
\label{sec:reasonable}

Our IGM parameter space, $(Q_{\rm HII}, \aveGammaHII)_{z7}$, is conservatively broad.  It would therefore be useful to estimate what evolution is ``physically reasonable'', in order to (qualitatively) tighten constraints.  We discuss this briefly below.

\subsubsection{Reionization}
\label{sec:reion_evo}

The evolution of the final stages of reionization depends on the evolution of the sources (galaxies) and sinks (Lyman limit systems, LLSs, and more diffuse systems) of ionizing photons.  The filling factor of HII regions depends on the source and sink terms as (e.g. \citealt{SG87}):
\begin{equation}
\label{eq:QHII}
\frac{d\QHII}{dt} = \frac{d}{dt} [f_\ast f_{\rm esc} f_{\rm b} N_\gamma f_{\rm coll}] - \alpha_B n_{\rm b} C (1+z)^3 \QHII ~ ,
\end{equation}
where $N_\gamma$ is the number of ionizing photons per stellar baryon, $f_{\rm esc}$ is the fraction of UV ionizing photons that escape into the IGM, $f_\ast$ is the fraction of galactic gas in stars, and $f_{\rm b}$ is the fraction of baryons inside the galaxy with respect to the cosmic mean $\Omega_{\rm b}/\Omega_{\rm m}$, $f_{\rm coll}$ is the fraction of mass collapsed into halos, $C$ is the gas clumping factor.  

Estimating this evolution is intrinsically complicated, as these parameters have a dependence on time, space, and (for the source terms) halo mass.  Usually, the simplification is made that $f_{\rm coll}$ evolves more rapidly than the other terms, and thus governs the evolution of reionization.  One can construct an extreme model (maximizing d$\QHII$/dt), by assuming that reionization is driven by only galaxies inside the most massive, rapidly-evolving halos.  If one uses $M_{\rm halo} \gsim 10^{11} \Msun$, roughly corresponding to the observed Lyman break galaxies (LBG) at high-redshift (e.g. \citealt{DF12}), the resulting maximal evolution over $\Delta z \approx 1$ is $\Delta \QHII \lsim 0.5$--0.6 (e.g. Fig. 1 in \citealt{Lidz07}).  
 However, this model would require a 'tuned', rapid fall of the '$f_{}$' parameters of the source term in eq. \ref{eq:QHII} at $z\sim6$ in order counter the rapid rise in d$f_{\rm coll}(\Mmin)$/dt, so as to reproduce the observed flat emissivity evolution at $z\lsim5$ (e.g. \citealt{Miralda-Escude03}; Fig. 12 in \citealt{MMS12}).

Furthermore, it is quite likely that the 'sinks' regulate the final stages of reionization.  As HII regions grow, they become 'recombination-limited' \citep{FO05}, and an increasing fraction of the ionizing photons goes into balancing recombinations.  In the late stages of reionization, this approximately translates to a rapid evolution in the clumping factor of the sink term in eq. \ref{eq:QHII}.  The most complete study to date was done by \citet{SM14}, who tracked the inhomogeneous evolution of sources and sinks via a sub-grid approach.  Their 'FULL' model results in an evolution of $\Delta \QHII \lsim 0.2$ over the final $\Delta z \approx 1$ of the EoR. 
However, it is likely that the late-stage evolution of $\QHII$ in \citet{SM14} is somewhat underestimated since: (i) they did not account for internal feedback, which can raise the value of $\Mmin$, speeding up d$f_{\rm coll}(\Mmin)$/dt; (ii) the analytic formula used in the sub-grid density distribution \citep{MHR00} is inaccurate at very high densities \citep{BB09} resulting in an underestimate of the emissivity evolution in the very late stages of (and following) reionization \citep{MOF11}.

From the above, we conclude that over $\Delta z \approx 1$, it is reasonable to assume $\Delta \QHII \lsim$ 0.2--0.5.  From the UV faint subsample in figures \ref{fig:Lya_frac} and \ref{fig:ratio_faint}, we see that this $\Delta \QHII$ requires $\aveGammaHII \lsim0.02$ to be consistent with $z=7$ data at 1$\sigma$.

\subsubsection{Photo-ionizing background}
\label{sec:gamma}

\begin{figure}
\vspace{-1\baselineskip}
{
\includegraphics[width=0.5\textwidth]{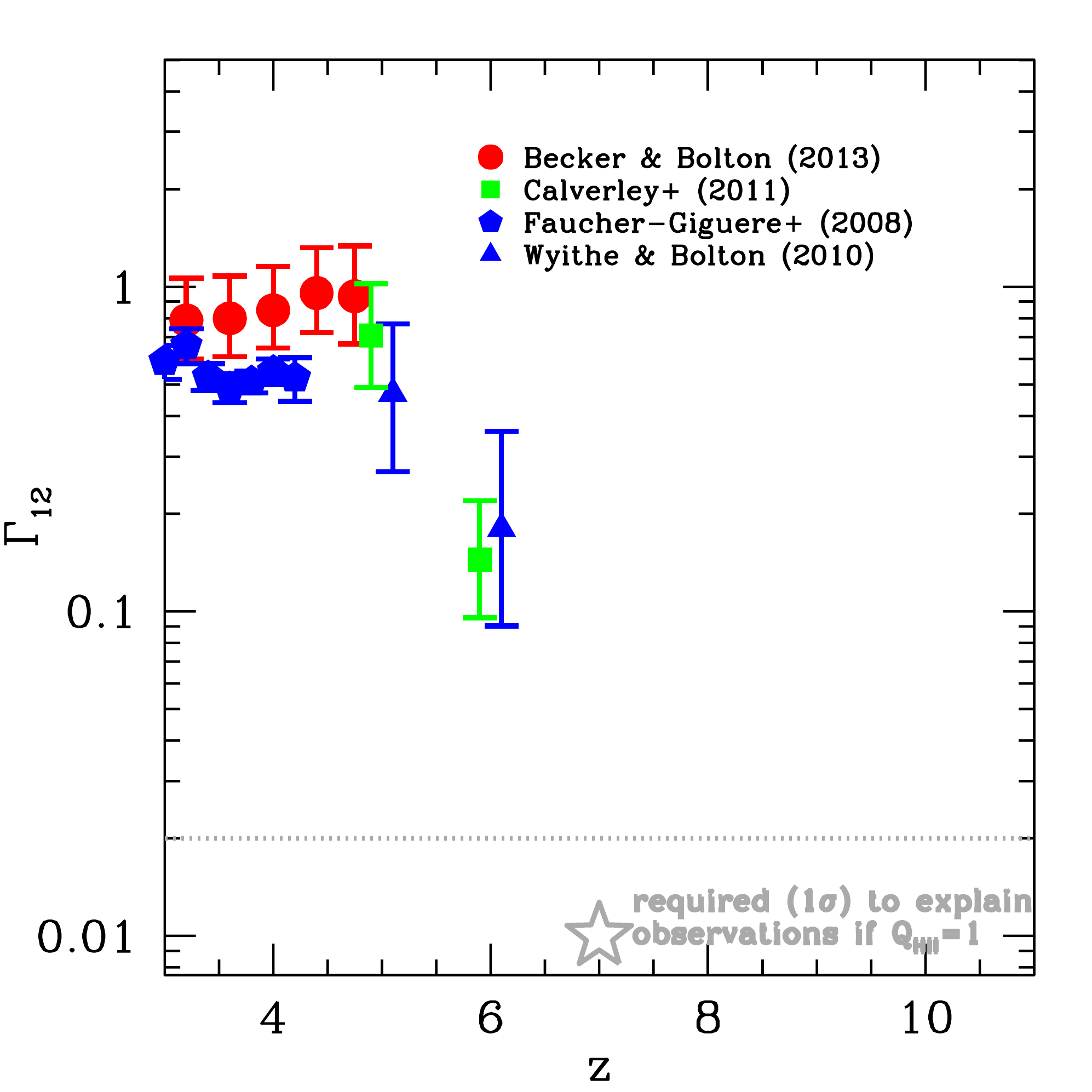}
}
\caption{
Estimates of the photo-ionizing background at $z \leq 6$ (in units of 10$^{-12}$ s$^{-1}$). Points at $z=5$, 6 are offset for clarity.  Circles correspond to \citet{BB13}, squares to \citet{Calverley11}, pentagons to \citet{Faucher-Giguere08}, and triangles to \citet{WB10}.  We also note with a gray star the maximum value of the photo-ionizing background at $z=7$ required to be consistent at 1-$\sigma$ with the \lya\ fraction, without evoking reionization. As seen in the text, the evolution in the \lya\ fraction is relatively insensitive to the photo-ionizing background for values of  $\gsim 0.02$.  This transition from the \lya\ fraction evolution being dominated by $Q_{\rm HII}$ to $\aveGammaHII$ is denoted by a horizontal dashed line.}
\label{fig:gamma_evo}
\vspace{-0.5\baselineskip}
\end{figure}

Can we say something about the value of $\aveGammaHII$ at $z=7$?  In Fig. \ref{fig:gamma_evo}, we show observational estimates of the photo-ionizing background at $z \leq 6$.  The data does show tentative evidence for a slight evolution over $z=5\rightarrow6$, though a much larger drop at $z=6\rightarrow7$ would be required to be consistent with the \lya\ fraction observations at 1$\sigma$, without evoking reionization (c.f. right axis of Fig. \ref{fig:ratio_faint}).

 From the theoretical side, any complete model of inhomogeneous reionization should also be able to self-consistently predict $\aveGammaHII$.  Indeed, \citet{SM14} predict a change of $\aveGammaHII$ by a factor of $\sim2$ over $\Delta z \approx 1$ in the final stages of reionization; however, as we mentioned above, the model of \citet{SM14} starts to break down post-reionization.  A more rapid evolution could be achieved if the photo-ionizing background is dominated by quasars at $z\sim7$ \citep{CF06}, though empirical models disfavor a strong contribution from quasars at $z\gsim4$ \citep{HM12}.

Further intuition can be gained if one simplifies the problem, studying the ionized IGM independently of the evolution of reionization.  Using small-box, high-resolution radiative transfer simulations of the post-reionization IGM, \citet{MOF11} show that a small change in the emissivity, $\epsilon$, could drive a rapid change in the photo-ionizing background, with $\Gamma \propto \epsilon^{3.5-4.5}$ at $z\sim6$.  In this case, the factor $\sim20$ drop in $\aveGammaHII$ from $z=6\rightarrow7$ required to explain the \lya\ fraction without reionization, would still require $\epsilon$ to drop by a factor of a $\sim$ few\footnote{The additional evolution of the scaling relation between $\Gamma$ and $\epsilon$ over this interval makes it difficult to present an accurate estimate.} likely dropping to below the value required to reionize the Universe at $z \sim 7$.  Hence, if one wishes to have such a low $\aveGammaHII(z=7)$ and have reionization occur earlier at $z>7$, the emissivity would likely have to evolve non-monotonically, with 
 an additional, seemingly ad-hoc population of ionizing sources at higher redshifts.  Hence we conclude that it is implausible for an evolution in self-shielded systems to fully explain the \lya\ fraction observations, without reionization, at the 1$\sigma$ level.


\section{Conclusions}
\label{sec:conc}

Motivated by recent observational claims of a rapid drop at $z>6$ in the \lya\ fraction, we construct a tiered model for simulating the IGM attenuation.  We combine large-scale semi-numeric reionization simulations with moderate-scale hydrodynamic simulations of the ionized IGM.  Thus for the first time we account for the opacity of {\it both} patchy reionization {\it and} proximate, high-column density systems inside HII regions.  

Using 5000 sightlines originating from $\sim10^{11}\Msun$ halos, we compute the average IGM transmission at $z=7$ in a 2D parameter space consisting of (i) the filling factor of ionized regions, $Q_{\rm HII}$; and (ii) the average photo-ionization rate inside HII regions, $\aveGammaHII$.  Our reionization morphologies are based on recent sub-grid models of UVB feedback on galaxies and an evolving recombination rate.  The resulting morphologies are characterized by smaller HII regions, making it easier for reionization to explain the \lya\ fraction evolution, compared to prior studies.

However, our models also include several, non-standard ingredients which act in the opposite direction, decreasing the predicted \lya\ fraction evolution from $z=6\rightarrow7$: (i) we use an intrinsic \lya\ emission profile which is offset from the galaxy's systemic redshift (instead of just computing the opacity at the systemic redshift); (ii) we include the opacities of self-shielded systems also when computing the $z=6$ transmission (instead of assuming zero opacity for all photons redward of the systemic redshift); and (iii) we use a calibrated, continuous prescription for self-shielding inside HII regions (instead of a step-function transition in the neutral fraction at a given density threshold).

Taken separately,
extreme  evolution over
 $z=6\rightarrow7$ in either reionization [$Q_{\rm HII}(z\approx7) \lsim$ 0.3] or the photo-ionizing background [$\aveGammaHII(z\approx7) < 0.01$] is required to be consistent with observations at the 1-$\sigma$ level.  Assuming a joint evolution of IGM properties allows somewhat more reasonable models to be consistent at 1$\sigma$ with data. Marginalizing over a uniform prior in $\log \aveGammaHII$, we obtain the constraint $Q_{\rm HII}(z\approx7) \lsim$ 0.6 (1-$\sigma$).
We caution however that all of our parameter space is consistent with observations at 2$\sigma$, motivating larger galaxy samples.

Given our sample of $z\approx6$ REWs, we predict that the \lya\ fraction cannot drop by more than a factor of $\sim2$ with IGM attenuation alone, even for ionized fractions as low as $Q_{\rm HII}\gsim$0.1. Larger evolutions would require a co-evolution with galaxy properties (e.g. \citealt{JSE12, Dijkstra14}).

We also note that the attenuation in our patchy reionization models has a unimodal distribution over various sightlines (as already noted in prior \lya\ emitter reionization studies; e.g. \citealt{McQuinn07LAE, MF08LAE}).  This is in contradiction to recent Bayesian analyses (e.g. \citealt{Treu12, Treu13, Pentericci14, Tilvi14}) which use an empirical 'on/off' model (dubbed 'patchy') as a proxy for inhomogeneous reionization.  In contrast, the attenuation from the {\it ionized} IGM, i.e. self-shielded systems, is relatively more an 'on/off' process: a LOS either encounters a nearby high-column density system and the \lya\ is strongly absorbed, or it does not and the attenuation is modest.  This behavior motivates the use of more physical models for the change in IGM transmission.

\vskip+0.3in

We thank J. Bolton for comments on a draft version of this manuscript.  We thank the Italian SuperComputing Resource Allocation (ISCRA) for computational resources (proposal LAE\_STAT).

\bibliographystyle{mn2e}
\bibliography{ms}

\end{document}